\documentclass[a4paper, amsfonts, amssymb, amsmath, reprint, showkeys, nofootinbib, twoside, superscriptaddress, longbibliography]{revtex4-2}
\usepackage[table]{xcolor} 
\usepackage[english]{babel}
\usepackage[utf8]{inputenc}
\usepackage[colorinlistoftodos, color=green!40, prependcaption]{todonotes}
\usepackage[pdftex, pdftitle={Article}, pdfauthor={Author}]{hyperref} 
\usepackage{amsthm}
\usepackage{mathtools}
\usepackage{physics}
\usepackage{graphicx}
\usepackage[left=18mm,right=18mm,top=25mm,columnsep=15pt, bottom=20mm]{geometry}
\usepackage{adjustbox}
\usepackage{placeins}
\usepackage[T1]{fontenc}
\usepackage{lipsum}
\usepackage{csquotes}
\usepackage{xfrac}
\usepackage{enumitem}
\usepackage{array}
\usepackage{multirow}
\usepackage{svg}

\renewcommand{\arraystretch}{1.6}

\begin{document}

\title{Multi-detector characterization of gravitational-wave burst tensor polarizations with the \textit{BayesWave} algorithm}

\author{Yi Shuen C. Lee}
    \email[]{ylee9@student.unimelb.edu.au}
    \affiliation{School of Physics, University of Melbourne, Parkville, Victoria 3010, Australia.}
    \affiliation{Australian Research Council Centre of Excellence for Gravitational Wave Discovery (OzGrav),
University of Melbourne, Parkville, Victoria 3010, Australia.} 
\author{Siddhant Doshi}
    \email[]{dsiddhant@gatech.edu}
    \affiliation{Center for Relativistic Astrophysics, Georgia Institute of Technology, Atlanta, GA 30332, USA.}
\author{Margaret Millhouse}
    \email[]{mmillhouse3@gatech.edu}
    \affiliation{School of Physics, University of Melbourne, Parkville, Victoria 3010, Australia.}
    \affiliation{Australian Research Council Centre of Excellence for Gravitational Wave Discovery (OzGrav),
University of Melbourne, Parkville, Victoria 3010, Australia.} 
    \affiliation{Center for Relativistic Astrophysics, Georgia Institute of Technology, Atlanta, GA 30332, USA.}
\author{Andrew Melatos}
    \email[]{amelatos@unimelb.edu.au}
    \affiliation{School of Physics, University of Melbourne, Parkville, Victoria 3010, Australia.}
    \affiliation{Australian Research Council Centre of Excellence for Gravitational Wave Discovery (OzGrav),
University of Melbourne, Parkville, Victoria 3010, Australia.}

\begin{abstract}

Einstein's theory of general relativity predicts that gravitational waves (GWs) are tensor-polarized, with two modes of polarization: plus ($h_+$) and cross ($h_\times$). The unmodeled GW burst analysis pipeline, \textit{BayesWave}, offers two tensor-polarized signal models: the elliptical polarization model ($E$) and the relaxed polarization model ($R$). Future expansion of the global GW detector network will enable more accurate studies of GW polarizations with GW bursts. Here a multi-detector analysis is conducted to compare the performance of $E$ and $R$ in characterizing elliptical and nonelliptical GW polarizations, using nonprecessing and precessing binary black holes (BBHs) respectively as representative synthetic sources. It is found that both models reconstruct the elliptical nonprecessing BBH signals accurately, but $E$ has a higher Bayesian evidence than $R$ as it is has fewer model parameters. The same is true for precessing BBHs that are reconstructed equally well by both models. However, for some events with high precession and especially with three or more detectors, the reconstruction accuracy and evidence of $R$ surpass $E$. The analysis is repeated for BBH events from the third LIGO-Virgo-KAGRA observing run, and the results show that $E$ is preferred over $R$ for existing detections. Although $E$ is generally preferred for its simplicity, it insists on elliptical polarizations, whereas $R$ can measure generic GW polarization content in terms of Stokes parameters. The accuracy of $R$ in recovering polarization content improves as the detector network expands, and the performance is independent of the GW signal morphology.

\end{abstract}

\maketitle

\section{Introduction}
\label{sec:intro}

 The global gravitational-wave (GW) detector network currently comprises four interferometers, namely the Hanford and Livingston Laser Interferometer Gravitational-Wave Observatory (LIGO) detectors in the USA \cite{LIGOmain}, the Virgo detector in Italy \cite{VIRGOmain} and the Kamioka Gravitational Wave Detector (KAGRA) in Japan \cite{KAGRA1}. With the commissioning of LIGO-India \cite{LIGOIndia} in progress and the proposals to build next-generation ground-based interfrometers like the Cosmic Explorer \cite{CE} and Einstein Telescope \cite{ET}, the global detector network will grow in the coming years. Benefits of a larger network include increased duty cycle, sky coverage, signal-to-noise ratio (SNR) and accuracy of source localization \cite{Schutz_Networks, LVK_localisation, Pankow_SkyLoc}.

Larger networks also enable more accurate studies of GW polarization structure. General relativity (GR) predicts that GWs are a linear combination of two transverse-traceless tensor polarizations, namely plus ($+$) and cross ($\times$). Metric theories beyond GR allow up to six independent polarizations, comprising additional vector modes ($x$ and $y$) and scalar modes ($b$ and $l$) \cite{Eardley1973_1,Eardley1973_2}. The overall amplitude of the metric perturbation in the time domain observed by a detector $I$ is given by \cite{Chatziioannou_2021}
\begin{equation}
    h_I(t) = \sum_P F^P_I(t,\psi,\Omega)h^P\left[t+\Delta t_I(\Omega)\right],  
    \label{eq:antenna_response}
\end{equation}
for $P\in\{+,\times, x, y, l, b\}$. In other words, the sensitivity of an individual GW detector to the amplitude $h^P$ of polarization mode $P$ is determined by a unique antenna pattern $F^P$, which is a function of time, the source orientation\footnote{$\psi$ is also known as the polarization angle. The standard LIGO-Virgo-KAGRA (LVK) definition of $\psi$ can be found in Refs. \cite{polarizationAngle_1, polarizationAngle_2}.} $\psi$ and the sky location $\Omega=\{\alpha, \delta\}$ expressed in terms of the right ascension ($\alpha$) and declination ($\delta$). $\Delta t_I(\Omega)$ is the arrival time delay between detector $I$ and a nominal reference location.

Long-duration (continuous) GW signals can be used to measure polarizations. In theory, this can be done with a single detector by searching for the unique $t$ dependence in $F^P_I(t,\psi,\Omega)$ \cite{CW1}. Multiple detectors improve polarization detection \cite{CW2}. Previous studies focus on persistent sources like a stochastic GW background  \cite{CW3, CW6, CW7, CW8, Tsukada_2023} and spinning neutron stars with an asymmetric moment of inertia \cite{CW4, CW5}. To date, however, these signals have not been detected by ground-based interferometric detectors. In contrast, LIGO \cite{LIGOmain} and Virgo \cite{VIRGOmain} have detected 90 compact binary coalescences (CBCs) up to the third observing run (O3) \cite{GWTC1, GWTC2, GWTC3}, and over 100 detection candidates from the ongoing fourth observing run (O4). Transient GW (burst) signals from CBCs typically last a few milliseconds to a few seconds, during which $F^P_I$ in Equation \ref{eq:antenna_response} is roughly constant. Since $h^P$ is the same for all detectors up to $\Delta t_I$, $n$ detectors are required at minimum to break $n$ polarization degeneracies if $\psi$ and $\Omega$ are well-constrained \cite{Takeda_GWPolCBC}. However, differential-arm detectors like LIGO, Virgo and KAGRA cannot distinguish between the two scalar modes ($l$ and $b$) because the antenna pattern functions are degenerate with respect to $l$ and $b$, meaning that only five polarization modes can effectively be disentangled with current detector designs~\cite{Isi_GWPol}. This is still enough in principle to distinguish between GR (tensor) and non-GR (non-tensor) polarizations. 

This paper focuses on characterizing GR-consistent (tensor) polarization structures. We employ the \textit{BayesWave} burst analysis algorithm \cite{BayesWave, BayesWave2, BayesWave3} to detect and characterize the GW bursts with minimal assumptions. \textit{BayesWave} provides two tensor-polarized signal models: the elliptical ($E$) and the relaxed ($R$) polarization models \cite{BayesWave, BayesWave2, BayesWave3}, represented as a sum of sine-Gaussian wavelets. $E$ restricts the frequency-domain amplitudes to $\tilde{h}_\times=i\epsilon \tilde{h}_+$ with ellipticity parameter $\epsilon$, whereas $R$ reconstructs $\tilde{h}_+$ and $\tilde{h}_\times$ independently. In this work, we present an empirical study on the polarization characterization performance of $E$ and $R$, for elliptical and nonelliptical GW burst signals. The Bayes factor between $E$ and $R$ is used to determine the model that best represents the data. According to GR, nonspinning binary black holes (BBHs) do not precess and emit elliptically polarized GWs \cite{Isi_GWPol}. Spinning CBCs, on the other hand, undergo precession if the component spin vectors orthogonal to the angular momentum vector are nonzero \cite{Schmidt_2015}. The angular momentum vector of precessing CBCs evolves with time, so the GW polarization is not strictly elliptical. Therefore we use simulated nonprecessing and precessing BBHs to represent the elliptical and nonelliptical GW burst sources respectively. We also quantify how accurately $R$ recovers GW polarization content of the nonprecessing and precessing BBHs in terms of Stokes parameters. This is not possible with $E$ because the overall polarization is described by a single ellipticity parameter $\epsilon$. The size of a detector network impacts \textit{BayesWave}'s signal characterization \cite{Lee_2021} and hence also the accuracy of measuring GW polarizations. For example, the Hanford (H) and Livingston (L) LIGO detectors are approximately coaligned, so their polarization measurements as in Equation \ref{eq:antenna_response} are not fully independent. As a result, the HL network provides limited information about the signal polarization structure. Thus we also compare \textit{BayesWave}'s polarization characterization performance between the HL (two-detector), HL-Virgo (three-detector) and HLV-KAGRA (four-detector) networks. 

The remaining sections of this paper are organized as follows. Section \ref{sec:BW} summarizes the \textit{BayesWave} algorithm and the signal models $E$ and $R$. Section \ref{sec:injections} details the nonprecessing and precessing BBH injection sets used in this study. Section \ref{sec:RvsE} discusses the aim and methods of comparing the burst characterization performance of $E$ and $R$; the results are presented in Section \ref{sec:BF_results} and the demonstration of methodology using real GW events from O3 is presented in Section \ref{sec:O3}. Section \ref{sec:GWpol} details the methods to quantify the accuracy of $R$ in recovering GW polarization; the multi-detector comparison for the nonprecessing and precessing BBHs is presented in Section \ref{sec:stokes_results}. We summarize our findings in Section \ref{sec:conclusion}.

\section{BayesWave and Signal Models}
\label{sec:BW}
\textit{BayesWave} is a source-agnostic analysis algorithm used by the LVK community to jointly characterize instrumental glitches and generic GW bursts \cite{GWTC1, GWTC2, GWTC3, allskyO1, allskyO2, allskyO3}. The algorithm uses a sum of sine-Gaussian wavelets to reconstruct transient and non-Gaussian features in the data, without prior assumptions about their source. A sine-Gaussian wavelet is parameterised by its central time $t_0$, central frequency $f_0$, quality factor $Q$, amplitude $\mathcal{A}$ and phase $\phi_0$, packaged into a vector $\vb*{\lambda}$. It is expressed mathematically in the time domain as
\begin{equation}
    \label{eq:wavelet}
    \Psi(t;\vb*{\lambda}) = \mathcal{A}e^{-(t-t_0)^2 / \tau^2}\cos\left[2\pi f_0(t-t_0)+\phi_0\right],
\end{equation}
with $\tau\equiv Q/(2\pi f_0)$. \textit{BayesWave} uses a trans-dimensional Reversible Jump Markov Chain Monte Carlo (RJMCMC) sampler to marginalize over the wavelet parameters and the number of wavelets (i.e. model complexity) needed to fit the data. Although RJMCMC is computationally expensive, it has enough flexibility to characterize generic GW bursts without assuming a physically motivated waveform a priori.

\textit{BayesWave} currently offers four independent phenomenological models: the Gaussian noise model ($\mathcal{N}$), the signal plus Gaussian noise model ($\mathcal{S}$), the glitch plus Gaussian noise model ($\mathcal{G}$), and the joint model ($\mathcal{S}$+$\mathcal{G}$) which simultaneously allows for a signal and a glitch. $\mathcal{S}$ assumes a coherent signal across all detectors in the network, whereas $\mathcal{G}$ is specific to each detector. Detailed descriptions of \textit{BayesWave}'s models can be found in Refs. \cite{BayesWave2, BayesWave3}. The model that best describes the data is determined by calculating the Bayes factor, i.e. the ratio of Bayesian evidences. \textit{BayesWave} uses thermodynamic integration to calculate the evidences (see Appendix \ref{sec:TI}).

In this paper, we focus on comparing two GR-consistent signal models with different polarization restrictions: elliptical ($E$) and relaxed ($R$), with $\{E, R\}\in \mathcal{S}$. We describe these models in Sections \ref{sec:elliptical_model} and \ref{sec:relaxed_model} respectively.

\subsection{Elliptical polarization, ${E}$}
\label{sec:elliptical_model}

Equation \ref{eq:antenna_response} specifies the time-domain response of a GW detector $I$ to a combination of polarization modes. In GR, the scalar and vector modes are omitted. As \textit{BayesWave} calculates the likelihoods in the frequency domain, we rewrite the detector response as
\begin{equation}
    \tilde{h}_I(f) = \left[ F^\times_I(\psi, \Omega) \tilde{h}^\times(f) + F^+_I(\psi, \Omega) \tilde{h}^+(f) \right]  e^{2\pi i f \Delta t_I(\Omega)}.
    \label{eq:det_response}
\end{equation}
\textit{BayesWave} analyzes burst signals much shorter than one day, so the diurnal and annual variations in $F^P_I(t,\psi,\Omega)$ are negligible, and we can write $F^P_I(t,\psi,\Omega)\approx F^P_I(\psi,\Omega)$. Equation \ref{eq:det_response} assumes that GWs travel at the speed of light and are not dispersed \cite{Isi_GWPolCBC, Chatziioannou_2021}. Hence GW signals are coherent across all detectors and can be represented by a single projected waveform. 

In order to disentangle $\tilde{h}_+(f)$ and $\tilde{h}_\times(f)$ from the measurements of $\tilde{h}_I(f)$ as in Equation \ref{eq:det_response}, one needs to know $\psi$ and $\Omega$, which are extrinsic parameters of \textit{BayesWave} signal model. In general, at least two detectors are required to break degeneracies between the two polarization modes, and additional detectors are required to constrain $\psi$ and $\Omega$. In the first observing run (O1) in 2015, only the two LIGO detectors were operating, whose arms are approximately parallel, leading to polarization degeneracies. Therefore the first signal model implemented in \textit{BayesWave} \cite{BayesWave} assumes elliptical polarization ($E$), where $\tilde{h}^+$ is expressed as a sum of $N$ wavelets and $\tilde{h}^\times$ is proportional to $\tilde{h}^+$ and out of phase by $\pi/2$ radians, viz. 
\begin{gather}
    \tilde{h}^+ = \sum_{n=1}^N \Psi\left[f; t^{(n)}_0, f^{(n)}_0, Q^{(n)}, \mathcal{A}^{(n)}, \phi^{(n)}_{0}\right],\label{eq:E_model_plus}\\
     \tilde{h}^\times=i\epsilon\tilde{h}^+.
    \label{eq:E_model_cross}
\end{gather}
The ellipticity parameter $\epsilon$ takes a value between zero (linear) and one (circular). Altogether $E$ is parameterized by $5N$ intrinsic parameters from the wavelets plus the four extrinsic parameters $\epsilon, \psi,$ and $\Omega=(\alpha, \delta)$.

\subsection{Relaxed polarization, ${R}$}
\label{sec:relaxed_model}

The model $E$ has achieved success in reconstructing GW signals like binary black holes (BBH) \cite{Ghonge_2020}, binary neutron stars (BNS) post-merger emissions \cite{BW_BNS_1, BW_BNS_2},  eccentric BBHs \cite{Dalya_2021}, and core-collapse supernovae (CCSNe) \cite{Raza_2022}. Most CBCs like BBHs and BNSs are elliptically polarized as in Equation \ref{eq:CBC_pol}, whereas generic bursts like CCSNe \cite{SN_pol1, SN_pol2} are not. Furthermore, $E$ does not hold for CBCs with time-dependent polarization content, e.g. signals with spin-precession and/or higher-order spherical harmonic modes (beyond $\ell=\abs{m}=2)$.

As Virgo and KAGRA join the global detector network, the task of separating the $+$ and $\times$ polarizations becomes more pressing. Each detector independently measures the combinations of polarizations as in Equation \ref{eq:det_response}, so expanded detector networks enable better separation of the polarization modes. In response, \textit{BayesWave} has been generalized to support the relaxed polarization model $R$ \cite{BayesWave3}:
\begin{align}
    \tilde{h}^+ &= \sum_{n=1}^N \Psi\left[f; t^{(n)}_0, f^{(n)}_0, Q^{(n)}, \mathcal{A}^{(n)}_+, \phi^{(n)}_{0,+}\right],  \label{eq:R_model_plus}\\
    \tilde{h}^\times&=\sum_{n=1}^N \Psi\left[f; t^{(n)}_0, f^{(n)}_0, Q^{(n)}, \mathcal{A}^{(n)}_\times, \phi^{(n)}_{0,\times}\right]. 
    \label{eq:R_model_cross}
\end{align}
The main difference between $E$ and $R$ is that $\tilde{h}^+$ and $\tilde{h}^\times$ are reconstructed separately in $R$ without the restriction imposed by Equation~\ref{eq:E_model_cross}. However, $\tilde{h}^+$ and $\tilde{h}^\times$ are not independent as they are both expressed as a sum of the same number ($N$) of wavelets, and the $n$-th plus and cross wavelets share the same central time $t_0^{(n)}$, central frequency $f_0^{(n)}$ and quality factor $Q^{(n)}$; only their amplitudes and phases differ. This arrangement is valid for two reasons. First, $\tilde{h}^+$ and $\tilde{h}^\times$ are generated by the same physical processes, so their time-frequency structure should be closely related~\cite{Chatziioannou_2021}. Second, according to Equation~\ref{eq:det_response}, the time-frequency structure of a coherent signal should be independent of the detector; only the amplitude and phase are modified when projected from the reference location onto the detector. With $R$, one can also set $\psi=0$ in Equation \ref{eq:det_response} without loss of generality\footnote{This is because $\psi$ and $\{\mathcal{A}^n_+,\mathcal{A}^n_\times\}$ are degenerate in Equation \ref{eq:det_response} i.e. the transformation $\{F^+(\psi, \Omega), F^\times(\psi, \Omega)\} \mapsto \{F^+(\psi', \Omega), F^\times(\psi', \Omega)\}$ for $\psi'=\psi+\Delta\psi$ can be fully described by $\{\mathcal{A}^{(n)}_+,\mathcal{A}^{(n)}_\times\}\mapsto \{\mathcal{A}^{(n)'}_+,\mathcal{A}^{(n)'}_\times\}$ (see Appendix A of Ref. \cite{CW2} for more details).}. Thus the overall dimensionality of $R$ is given by the $7N$ intrinsic wavelet parameters $t^{(n)}_0, f^{(n)}_0, Q^{(n)}, \mathcal{A}^{(n)}_+, \phi^{(n)}_{0,+}, \mathcal{A}^{(n)}_\times, \phi^{(n)}_{0,\times}$ plus two extrinsic parameters $\Omega=(\alpha, \delta)$.

\section{Injection datasets}
\label{sec:injections}

The two key goals of this paper are (i) to test whether model $E$ or model $R$ reconstructs more accurately the signals from nonprecessing (zero-spin) and precessing (nonzero-spin) BBHs, and (ii) to quantify how accurately $R$ recovers generic BBH polarizations in terms of Stokes parameters. We address these goals quantitatively in Sections \ref{sec:RvsE} and \ref{sec:GWpol} by analyzing two separate sets of simulated injections, one comprising nonprecessing BBHs and the other comprising precessing BBHs. We specify the two injection sets in this section. A previous study showed that \textit{BayesWave}'s detection statistic, viz. the log signal-versus-glitch Bayes factor, scales with the number of detectors in the network \cite{Lee_2021}. Furthermore, the size of the detector network affects the distinguishability of polarization modes and hence the accuracy of polarization measurements \cite{Takeda_GWPolCBC, Isi_GWPolCBC}. So we inject the BBHs into simulated noise of three detector configurations, namely Hanford-Livingston (HL, two detectors), Hanford-Livingston-Virgo (HLV, three detectors) and Hanford-Livingston-KAGRA-Virgo (HLKV, four detectors) to compare their performances in characterizing GW burst polarizations.

The nonprecessing and precessing BBH injection sets are simulated using the \texttt{IMRPhenomXPHM} waveform approximant \cite{XPHM1, XPHM2, XPHM3, XPHM4}. Each injection set consists of 200 waveforms with sky locations uniformly distributed on a sphere. The main difference between the two injection sets is the distributions of the dimensionless spin $\abs{\vb*{\chi}_{\mathcal{C}}}$, where the subscript $\mathcal{C}=1,2$ labels the primary and secondary black holes respectively. The effective precession spin parameter $0\leq\chi_p\leq1$ is conventionally used to characterize the degree of precession in BBHs \cite{Schmidt_2015}. Since $\chi_p$ depends on $\abs{\vb*{\chi}_{\mathcal{C}}}$ (see Appendix \ref{sec:chi_p}), we set $\abs{\vb*{\chi}_{\mathcal{C}}}=0$ to obtain $\chi_p=0$ for all of the nonprecessing BBH injections. For the precessing BBHs, $\vb*{\chi}_{\mathcal{C}}$ and $\chi_p$ evolve with time. Although we cannot guarantee precession throughout the coalescence because the motion is too complicated to predict analytically, we can ensure that the system precesses initially. This is achieved by sampling uniformly in the domain $0.1 \leq \abs{\vb*{\chi}_{\mathcal{C}}} \leq 1$ to obtain $\chi_{p,\rm init}\neq0$ initially at $f=16$ Hz for all precessing BBH injections. Equation \ref{eq:chi_p} shows that $\chi_p$ depends only on $\abs{\vb*{\chi}_{\mathcal{C}, \perp}}$ i.e. spin components orthogonal to the angular momentum vector of the binary system. In other words, only part of $\abs{\vb*{\chi}_{\mathcal{C}}}$ contributes to $\chi_p$. Therefore we impose a lower bound of 0.1 on $\abs{\vb*{\chi}_{\mathcal{C}}}$ to avoid obtaining injections with negligible $\abs{\vb*{\chi}_{\mathcal{C}, \perp}}$, and hence $\chi_{p,\rm init}$, which are indistinguishable from nonprecessing BBHs.

In this work, we use the nonprecessing (precessing) BBHs injections to compare the performance between $R$ and $E$ for elliptical (nonelliptical) GW signals. Therefore we need to ensure that the injection sets fulfill their respective polarization requirements. The polarization of nonprecessing CBCs is quantified by the ellipticity $\epsilon={\abs{h_\times(t)}}/{\abs{h_+(t)}}$, which depends on the inclination angle $\iota$ between the line of sight and the normal to the orbital plane \cite{Schutz_2022}, according to
\begin{equation}
    \frac{\abs{h_\times(t)}}{\abs{h_+(t)}}=\frac{2\cos\iota}{1+\cos^2 \iota}.
\label{eq:CBC_pol}
\end{equation}
Therefore, in theory, all inclination angles except $\iota = 90^\circ$ produce elliptically-polarized nonprecessing BBHs that are valid for our study. However, the source orientation affects the radiated power, e.g. $\iota = 90^\circ$ (edge-on) results in minimum power, and $\iota = 0^\circ$ or $180^\circ$ (face-on or face-off) results in maximum power~\cite{Schutz_Networks}. This, in turn, can affect the distinguishability of the GW polarization modes, so we select $\iota$ with caution. For nonprecessing BBHs, $\iota$ is fixed throughout the inspiral. In contrast, for precessing BBHs, $\iota$ evolves throughout the inspiral and the polarizations are not related as in Equation~\ref{eq:CBC_pol}. Since $\abs{\vb*{\chi}_{\mathcal{C}}}$ of the precessing BBHs is uniformly sampled, some injections are expected to undergo strong precession, leading to significant deviations from the initial $\iota$, while others will exhibit minimal precession, maintaining an inclination close to the initial $\iota$. Although only the strongly precessing BBHs are useful for comparing the $E$ and $R$ performance between nonprecessing and precessing signals, the minimally precessing BBHs can be used to assess the consistency of $E$ and $R$ compared to the nonprecessing BBHs, provided their $\iota$ values are similar. For this reason, we nominally set $\iota=45^\circ$ for all nonprecessing BBHs and initialize all precessing BBH inspirals with $\iota=45^\circ$ at $f=16$ Hz to facilitate the consistency check. The initial $\iota$ does not affect the strongly precessing BBHs, as they are expected to deviate from it regardless.

\begin{figure}
    \centering
    \includegraphics[width=.49\textwidth]{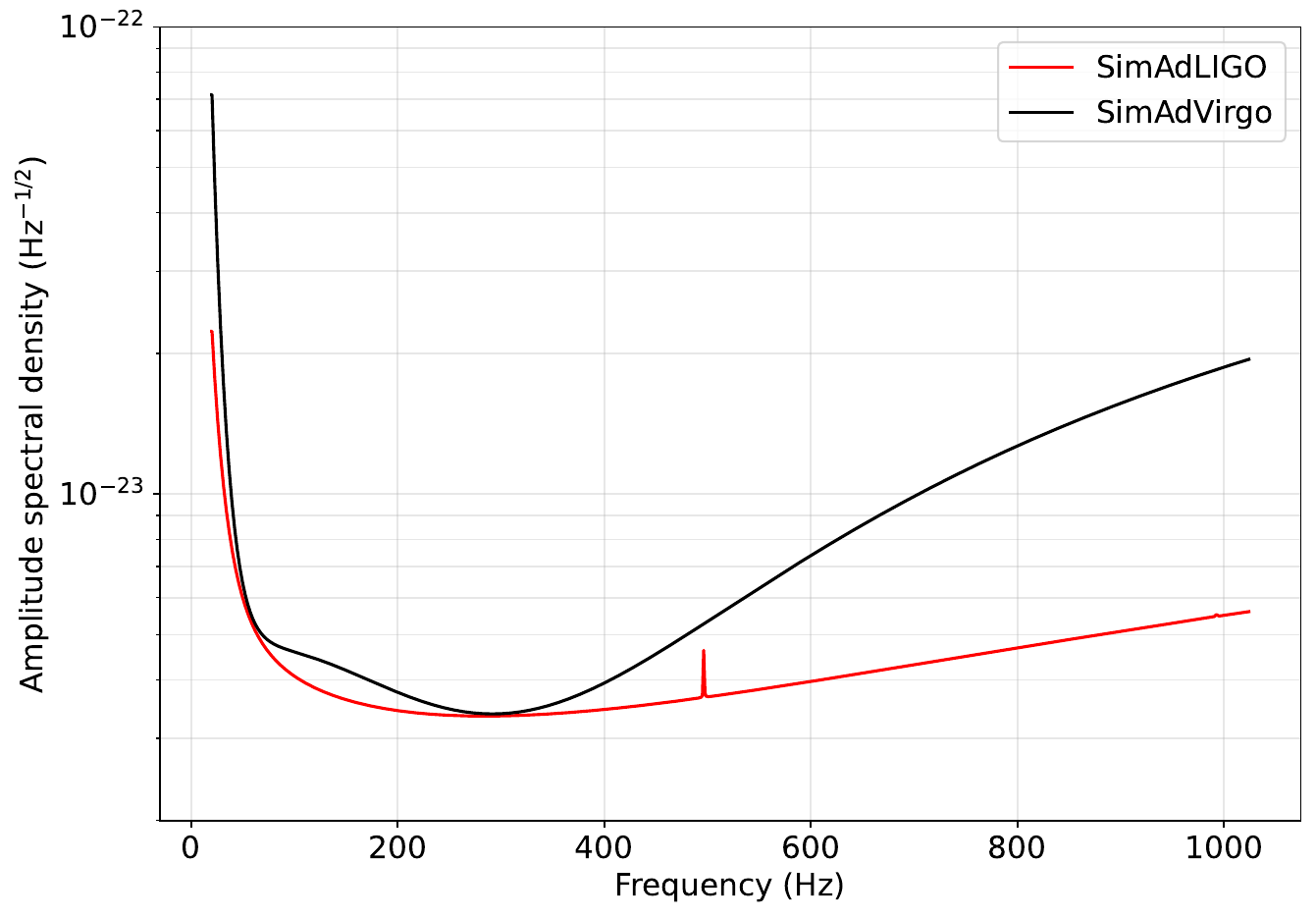}
    \caption{Simulated amplitude spectral density (ASD) curves of the Advanced LIGO (SimAdLIGO, red) and Virgo (SimAdVirgo, black) detectors from the LVK Algorithm Library \cite{lalsuite}. SimAdLIGO represents the LIGO Hanford (H) and Livingston (L) detectors, whereas SimAdVirgo represents Virgo (V) and KAGRA (K) in this study.}
    \label{fig:PSD}
\end{figure}

To compare $R$ versus $E$ fairly, we keep all other parameters fixed across both injection sets. Asymmetric masses are known to amplify precession-related GR effects \cite{Pratten_2020}. Hence we set the primary and secondary black hole masses to $m_1=40M_\odot$ and $m_2=8M_\odot$ respectively, to achieve a mass ratio of $q=m_1/m_2=5$ for all injections. The source distances are chosen to obtain an injected network signal-to-noise ratio $\rm SNR_{\rm net}=50$ in simulated HLV data. The same distance distribution is used for the HL and HLKV injections, i.e. the sources have the same inherent loudness, but $\rm SNR_{\rm net}$ is adjusted by removing Virgo contributions and adding KAGRA contributions respectively. As a result, the HLKV injections have the highest $\rm SNR_{\rm net}$, followed by HLV and then HL. All nonprecessing and precessing BBHs are injected into Gaussian noise colored by the sensitivity curves shown in Figure \ref{fig:PSD}. In this paper, with an eye to the future, we use the same sensitivity curves for Virgo and KAGRA. We also note that $\rm SNR_{\rm net}{\rm(HLV)}=50$ is a few factors higher than the BBH detections reported in the Gravitational-wave Transient Catalogs (GWTCs), where the HLV network matched-filter SNRs is typically $\leq25$ \cite{GWTC1, GWTC2, GWTC3}. Therefore the studies presented in this paper are a proof of concept. We do not use the results to make claims about the polarization properties of existing detections. 

 For each injection set, we perform two separate \textit{BayesWave} analyses with the signal models $E$ and $R$.  To ensure comparability between the two analyses, both analyses are configured the same. Firstly we assume fixed detector noise curves as in Figure \ref{fig:PSD} for all the analyses. By convention, we set the lower bound of the analysis frequency range to 20 Hz, to avoid the high-amplitude noise floor $<20$ Hz \cite{LSC_detNoise}. We choose a sampling rate of $2048$ Hz, where the Nyquist frequency sets the upper analysis frequency bound at $1024$ Hz. The overall analysis frequency range is therefore $20$-$1024$ Hz which is sufficient for the detection of BBHs. The analyzed segments are set to 4 seconds, which is longer than typical CBC signals ($\sim1$ second), to prevent truncating detectable segments of the BBH signals. 

\section{$R$ versus $E$}
\label{sec:RvsE}

Figure 1 of Ref. \cite{BayesWave3} qualitatively demonstrates that $R$ reconstructs BBH signals with spin-induced precession more accurately than $E$, assuming a zero-noise realization. The figure shows the injected and reconstructed time-domain signal amplitudes for one nonprecessing and two precessing BBHs. Both $E$ and $R$ perform comparably when reconstructing the nonprecessing BBH signal. On the other hand, the amplitudes of the precessing BBHs are modulated, and $E$ can reproduce the modulations to some extent, although they are represented better by $R$. However, better reconstruction does not necessarily imply that $R$ is supported more strongly by the data, once the Occam penalty in the Bayesian evidence is taken into account. Furthermore, the true signal morphology and hence the accuracy of reconstruction are unknown when analyzing real, astronomical data. 

In this work, we use the Bayes factor between $R$ and $E$ ($\ln\mathcal{B}_{R, E}$) to study whether there is a preferred polarization model for different signal morphologies. If there is, can $\ln\mathcal{B}_{R, E}$ be used to distinguish between elliptical and nonelliptical polarizations? We use the nonprecessing (elliptical) and precessing (nonelliptical) BBH injections described in Section \ref{sec:injections} as representative case studies. In Section \ref{sec:fom}, we define and justify the metrics used to compare $E$ and $R$. The results involving synthetic data are presented in Section \ref{sec:BF_results}. We also demonstrate how one can compute the relevant comparison metrics for real O3 events in Section \ref{sec:O3}.

\subsection{Comparison metrics}
\label{sec:fom}

\begin{figure*}
    \centering
    \includegraphics[width=\textwidth]{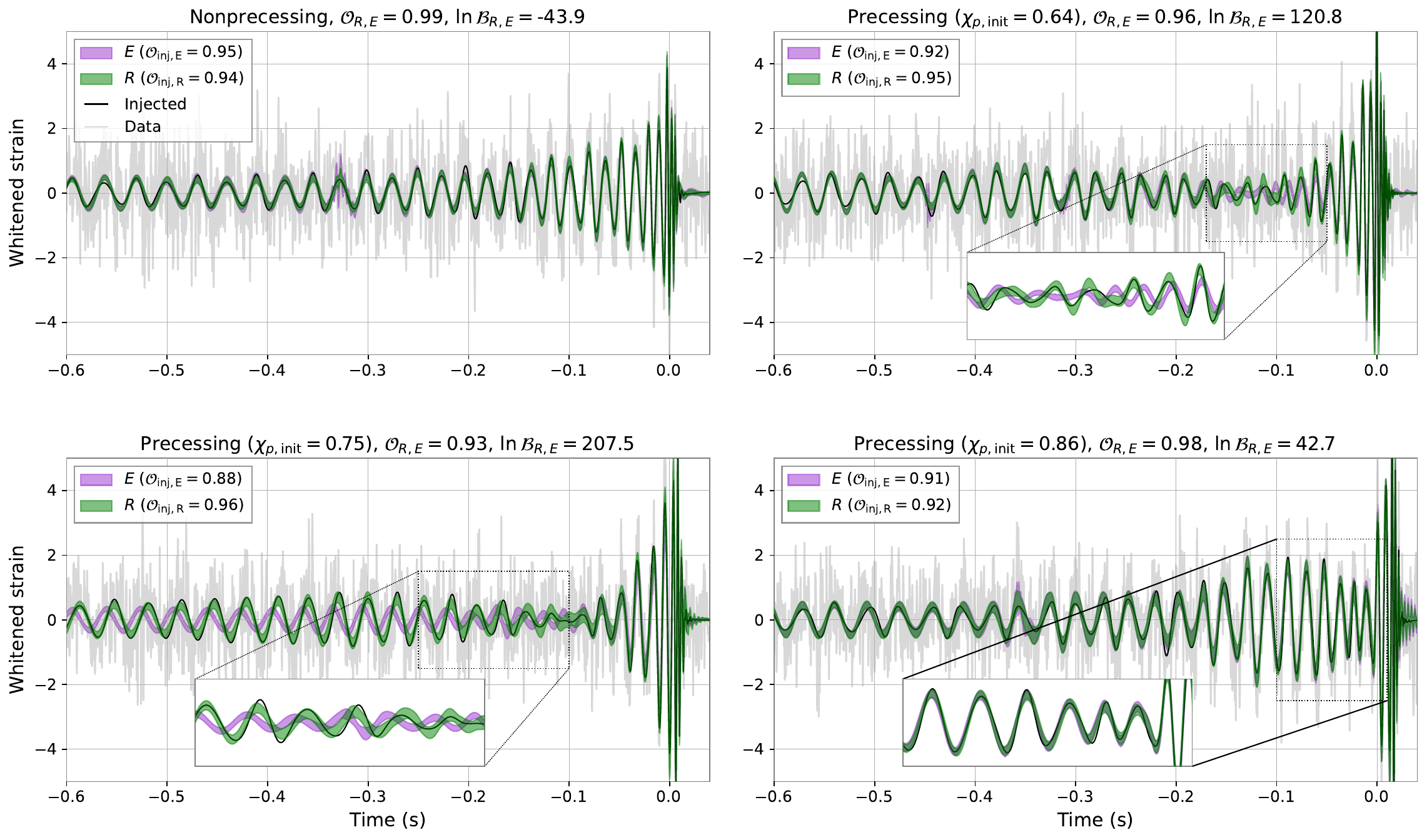}
    \caption{Injected and recovered time-domain signals for an arbitrarily chosen detector (LIGO Hanford). The top left panel shows the signal for a nonprecessing BBH; the remaining three panels show signals for precessing BBHs with $\chi_{p, \rm init}=0.64, 0.75$ and $0.86$. The black solid curves indicate the injected signal and the grey curve indicate the detector data, which includes both the signal and noise. The 90\% credible intervals for the waveforms recovered by $E$ and $R$ are indicated by the purple and green shaded bands respectively. $\mathcal{O}_{R,E}$ and $\ln\mathcal{B}_{R,E}$ for each event is shown at the top of the respective panels and the network overlaps $\mathcal{O}_{{\rm inj}, R}$ and $\mathcal{O}_{{\rm inj}, E}$ are shown in the legends. The overlaps displayed here are obtained from the HLKV analyses; these four events are injected into HLKV with $\mathrm{SNR}_{\mathrm{net}} \approx 65$.}
    \label{fig:sample_overlaps}
\end{figure*}

It is convenient to compute the log Bayes factor (i.e. the log evidence ratio) between $R$ and $E$,
\begin{equation}
    \ln\mathcal{B}_{R, E}=\ln p(\mathbf{d}|R) - \ln p(\mathbf{d}|E),
    \label{eq:B_RE}
\end{equation}
as a basis for model selection. $\ln\mathcal{B}_{R, E}>0$ indicates that $R$ is more strongly supported by the data $\mathbf{d}$ than $E$, and vice versa. The comparison via Equation \ref{eq:B_RE} is valid when the $E$ and $R$ analyses use the same noise power spectral density (PSD). That is, we assume the PSD is known and is the same in each analysis.

Our goal is to study how $\ln\mathcal{B}_{R, E}$ varies as the discrepancy increases between the signal waveforms reconstructed by $E$ and $R$. We apply the standard LVK definition of the network overlap between the signal waveforms $h^R$ and $h^E$ reconstructed by $R$ and $E$ respectively \cite{Ghonge_2020}:
\begin{equation}
    \mathcal{O}_{R, E} \equiv \frac{\langle {h}^R | {h}^E \rangle}{\sqrt{\langle {h}^R | {h}^R \rangle \langle {h}^E | {h}^E \rangle}}.
    \label{eq:overlap_RE}
\end{equation}
For a network with $\mathcal{I}$ detectors, one has
\begin{equation}
    \langle {h}^a | {h}^b \rangle = \sum_{i=1}^\mathcal{I}\langle \tilde{h}_i^a | \tilde{h}_i^b\rangle,
    \label{eq:network_innerProd}
\end{equation}
and the noise-weighted inner product between waveforms $\tilde{h}_i^a$ and $\tilde{h}_i^b$ of the $i$-th detector is given by
\begin{equation}
    \langle \tilde{h}_i^a | \tilde{h}_i^b \rangle = 4\Re \int_0 ^\infty df \, \frac{\tilde{h}^a_i (f) \tilde{h}^{b*}_i(f)}{S_{n,i}(f)}.
    \label{eq:noise_weighted_inner_prod}
\end{equation}
In Equation \ref{eq:noise_weighted_inner_prod}, $\tilde{h}^*$ denotes the complex conjugate of $\tilde{h}$ and $S_{n,i}(f)$ is the one-sided\footnote{It makes sense to work with the one-sided PSD $S_{n,i}(f)$, because the GW signal and detector output are real, which implies $S_{n,i}(-f)=S_{n,i}(f)$. This is also why the Fourier integral in Equation \ref{eq:noise_weighted_inner_prod} only includes positive frequencies \cite{GW_sensitivity_curves}.} noise PSD of detector $i$. By definition, $\mathcal{O}_{R, E}$ takes values between $-1$ and $1$, where $\mathcal{O}_{R, E}=1$ indicates perfect agreement between the $E$ and $R$ waveforms, $\mathcal{O}_{R, E}=0$ indicates no agreement and $\mathcal{O}_{R, E}=-1$ indicates anti-correlation.

For simulated injections, network overlaps between the injected and recovered signal waveforms ($\mathcal{O}_{{\rm inj}, R}$ and $\mathcal{O}_{{\rm inj}, E}$) are typically used to quantify the reconstruction accuracy of \textit{BayesWave}'s signal models i.e. they are good performance indicators. Thus we take advantage of the simulated injections to supplement our understanding about how $\mathcal{O}_{{\rm inj}, R}$ and $\mathcal{O}_{{\rm inj}, E}$ correlate with $\mathcal{O}_{R,E}$, for different BBH waveforms. We achieve this by evaluating the overlap ratio $\mathcal{O}_{{\rm inj}, R}/\mathcal{O}_{{\rm inj}, E}$ as a function of $\mathcal{O}_{R,E}$, where $\mathcal{O}_{{\rm inj}, R}/\mathcal{O}_{{\rm inj}, E}>1$ indicates that $R$ recovers the injected signal more accurately than $E$. However, $\mathcal{O}_{{\rm inj}, R}/\mathcal{O}_{{\rm inj}, E}$ quantifies the relative reconstruction accuracy of the two models. Hence we also present the results for $\mathcal{O}_{{\rm inj}, R}$ versus $\mathcal{O}_{{\rm inj}, E}$, to compare the absolute reconstruction accuracy of each model.

To illustrate how the overlaps ($\mathcal{O}_{R, E}$, $\mathcal{O}_{{\rm inj}, R}$, $\mathcal{O}_{{\rm inj}, E}$) quantify agreement between waveforms, we show the time-domain signal reconstructions of one nonprecessing BBH and three precessing BBHs with contrasting $\chi_{p, \rm init}$ in Figure \ref{fig:sample_overlaps}. The purple and green shaded regions show the 90\% credible interval of the $E$ and $R$ recovered waveforms respectively. Most of the purple and green shaded regions enclose the injected waveform (black solid curves), implying that both $E$ and $R$ reconstructions are comparable and consistent with the true waveform. This is also reflected in the high overlaps, i.e. $\mathcal{O}_{{\rm inj}, E}\geq0.88$ and $\mathcal{O}_{{\rm inj}, R}\geq 0.90$ as quoted in the legends. However, the two precessing events with $\chi_{p, \rm init}=0.64$ (top right) and $\chi_{p, \rm init}=0.75$ (bottom left) in Figure \ref{fig:sample_overlaps} show qualitatively better reconstructions with $R$ (green) then $E$ (purple). The insets show that the improvements with $R$ are especially prominent where the amplitudes are modulated. Quantitatively, we also obtain $\mathcal{O}_{{\rm inj}, R}>\mathcal{O}_{{\rm inj}, E}$ for these two events. Due to the discrepancies between $R$ and $E$, $\mathcal{O}_{R, E}$ is also comparatively lower (0.96 and 0.93) than for the other two events (0.98 and 0.99). We also find that $\ln\mathcal{B}_{R,E}$ for these two events (120.8 and 207.5) exceeds the other two ($-$43.9 for nonprecessing and 42.7 for precessing with $\chi_{p, \rm init}=0.86$). Altogether $\mathcal{O}_{R, E}$ and $\ln\mathcal{B}_{R,E}$ suggest that the $R$ model is strongly preferred over $E$ for the two precessing events with $\chi_{p, \rm init}=0.64$ and $\chi_{p, \rm init}=0.75$. Notably, $\ln\mathcal{B}_{R,E}$ for the event with $\chi_{p, \rm init}=0.86$ is lowest among the three precessing BBH examples, despite having the highest $\chi_{p, \rm init}$. Furthermore, its signal exhibits modulations like the other two precessing events, but the inset shows that the modulations are reconstructed equally well by $E$ and $R$, yielding $\mathcal{O}_{{\rm inj}, R}\approx\mathcal{O}_{{\rm inj}, E}$. 

To justify the above observations, we analyze the injected polarization content of the three precessing BBHs (plots omitted for brevity). We find that the $\chi_{p, \rm init}=0.86$ event has higher circular polarization than the other two precessing events. Therefore, $E$ characterizes the $\chi_{p, \rm init}=0.86$ event with an accuracy comparable to $R$. In summary, the above analysis implies two things. (1) Precessing BBHs evolve towards different inclination angles (and hence ellitpical polarizations) unpredictably, without a simple connection to $\chi_{p,{\rm init}}$. This is because it is possible for some BBHs with high initial precession to evolve into a stable configuration with slowly varying inclination before merging. Therefore, some BBHs with high $\chi_{p, \rm init}$ may emit more elliptical GW signals in the later stages of the inspiral. Conversely, BBHs with low $\chi_{p, \rm init}$ sometimes exhibit the opposite behavior. (2) Amplitude modulations in $h_+(t)$ and $h_\times(t)$ do not necessarily imply significant deviations from elliptical polarization. As a result, $E$ is able to reconstruct some of the modulations, but $R$ does better in general. These features agree with Figure 1 of Ref. \cite{BayesWave3}. 

\subsection{Nonprecessing versus precessing BBHs}
\label{sec:BF_results}

\begin{figure*}
    \begin{minipage}{\textwidth}
        \centering
        \includegraphics[width=\linewidth]{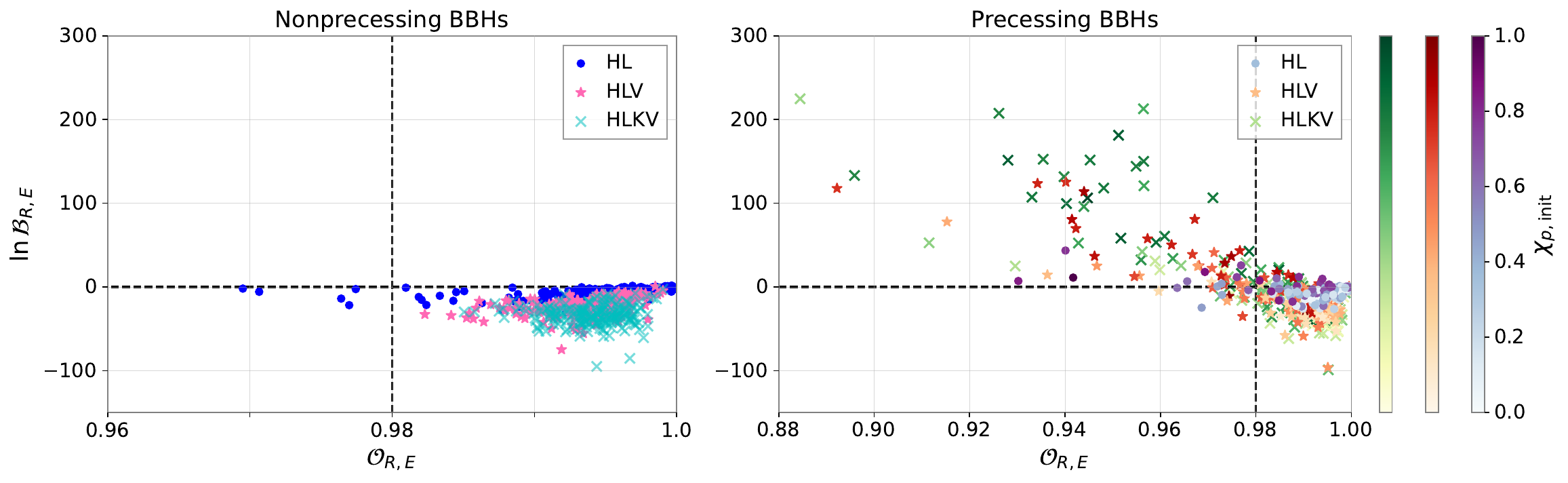}
    \end{minipage}\hfill
       \begin{minipage}{\textwidth}
        \centering
        \includegraphics[width=\linewidth]{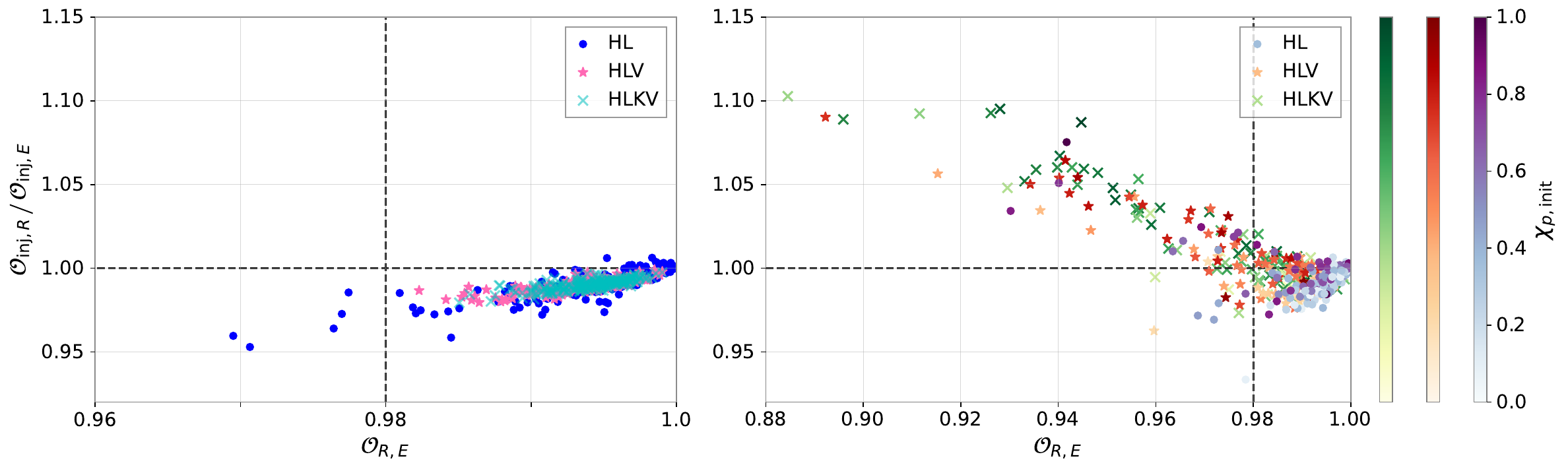}
    \end{minipage}\hfill

       \caption{Performance of $R$ versus $E$. The top row shows $\ln\mathcal{B}_{R, E}$ versus $\mathcal{O}_{R,E}$ and the bottom row shows $\mathcal{O}_{{\rm inj}, R}/ \mathcal{O}_{{\rm inj}, E}$ versus $\mathcal{O}_{R,E}$. In each row, the left panel shows the properties of the nonprecessing BBH injections for HL (dark blue circles), HLV (pink stars) and HLKV (light blue crosses). The right panel shows the properties of the precessing BBH injections for HL (purple circles), HLV (orange stars) and HLKV (green crosses). The horizontal scales are different for the left and right panels, but the same in the corresponding top and bottom panels. The three color bars per row apply to the right panels only; they indicate $\chi_{p,\rm init}$ of the precessing BBHs for HL, HLV, and HLKV in matching colors. All nonprecessing binaries in the left panels have $\chi_p=0$. The dashed lines in the top row at $\ln\mathcal{B}_{R, E}=0$ (horizontal) and $\mathcal{O}_{R,E}=0.98$ (vertical) divide the plot into four quadrants, as do the dashed lines at $\mathcal{O}_{{\rm inj}, R}/ \mathcal{O}_{{\rm inj}, E}=1$ (horizontal) and $\mathcal{O}_{R,E}=0.98$ (vertical) in the bottom row.}
    \label{fig:bayesfactor}
\end{figure*}

Figure \ref{fig:bayesfactor} presents the $R$ versus $E$ analyses for the nonprecessing and precessing BBH injections, using the comparison metrics described above. The top row plots $\ln\mathcal{B}_{R, E}$ as a function of $\mathcal{O}_{R, E}$; the bottom row plots $\mathcal{O}_{{\rm inj}, R}/ \mathcal{O}_{{\rm inj}, E}$ as a function of $\mathcal{O}_{R,E}$. All plots show results from the HL, HLV and HLKV analyses. The comparison of $R$ and $E$ through $\ln\mathcal{B}_{R, E}$ is meaningful, when the event is consistent with the signal model. Therefore Figure \ref{fig:bayesfactor} only plots events with signal evidences greater than the glitch and/or Gaussian noise model in the analysis, i.e. events that satisfy $\ln\mathcal{B}_{R, \mathcal{G}}>0$ and $\ln\mathcal{B}_{R, \mathcal{N}}>0$, and similarly for $E$. Less than three of the 200 injections are removed from the analysis per detector configuration for failing to satisfy these conditions. This is expected, because all BBHs in this study are injected with high SNR, so noise- or glitch-like events are scarce. All the removed events have low SNR in one or more detectors, resulting in higher evidence for the incoherent $\mathcal{G}$ model than the coherent signal models $R$ and $E$.

We first discuss the plots of $\ln\mathcal{B}_{R, E}$ versus $\mathcal{O}_{R, E}$ in the top row of Figure \ref{fig:bayesfactor}. In order to compare how the nonprecessing and precessing BBHs are distributed, we segment both plots in the top row using black vertical and horizontal dashed lines placed at $\mathcal{O}_{R, E}=0.98$ and $\ln\mathcal{B}_{R, E}=0$ respectively. We set $\mathcal{O}_{R, E}=0.98$ as the nominal threshold above which $E$ and $R$ models do equally well in fitting the data\footnote{We set the threshold as $\mathcal{O}_{R, E}=0.98$ instead of $\mathcal{O}_{R, E}=1$, because the model parameters of $E$ and $R$ differ, so it is unrealistic to expect their reconstructions to match exactly.}. $\ln\mathcal{B}_{R, E}=0$, by definition, is where the model evidences of $E$ and $R$ are equal i.e. neither model is preferred over the other from a Bayesian perspective. The top left plot in Figure \ref{fig:bayesfactor} shows that nonprecessing BBHs are generally found in the bottom right segment, with $\mathcal{O}_{R, E}>0.98$ and $\ln\mathcal{B}_{R, E}<0$. This is true for all three detector configurations. The evidence is higher for $E$ than $R$ because of the Occam penalty on unnecessary model dimensions of $R$. Previous studies have shown that for high SNR ($\gtrsim 10$) events, the Bayesian evidence of model $\mathcal{H}$ can be approximated by \cite{BayesWave2, Lee_2021}
\begin{equation}
    p(\mathbf{d}|\mathcal{H}) \simeq p(\mathbf{d}|\vb*{\theta}_{\rm MAP}, \mathcal{H})\frac{\Delta V_{\mathcal{H}}}{V_{\mathcal{H}}},
    \label{eq:evidence_appx}
\end{equation}
where $p(\mathbf{d}|\vb*{\theta}_{\rm MAP}, \mathcal{H})$ is the likelihood evaluated at the maximum \textit{a posteriori} (MAP) parameter value\footnote{$\vb*{\theta}=\vb*{\theta}_{\rm MAP}$ is the parameter value at which the posterior $p(\vb*{\theta}|\vb{d}, \mathcal{H})$ peaks.} $\vb*{\theta}_{\rm MAP}$, and $\Delta V_{\mathcal{H}}$ and $V_{\mathcal{H}}$ are respectively the posterior and prior volumes of $\mathcal{H}$. The fraction $\Delta V_{\mathcal{H}}/V_{\mathcal{H}}$ is known as the Occam penalty because it suppresses the evidence of models with unnecessarily large prior volume. Equations \ref{eq:B_RE} and \ref{eq:evidence_appx} can be combined to give
\begin{equation}
    \ln\mathcal{B}_{R, E}\simeq\ln\frac{p(\mathbf{d}|\vb*{\theta}_{\rm MAP}, R)}{p(\mathbf{d}|\vb*{\theta}_{\rm MAP}, E)} + \ln \frac{\Delta V_{R}}{\Delta V_{E}} + \ln \frac{V_{E}}{V_{R}}.
    \label{eq:BRE_appx}
\end{equation}
We find empirically $\mathcal{O}_{R, E}>0.98\approx1$ for the nonprecessing BBHs. The high $\mathcal{O}_{R, E}$ implies that the likelihoods of $E$ and $R$ are similar. As a result, we have $p(\mathbf{d}|\vb*{\theta}_{\rm MAP}, R)\approx p(\mathbf{d}|\vb*{\theta}_{\rm MAP}, E)$. Thus, the first term on the right-hand side of Equation \ref{eq:BRE_appx} is approximately zero. Likewise the second term is approximately zero, because $E$ and $R$ have approximately equal posterior volumes. Therefore when the reconstructed waveforms of $R$ and $E$ agree closely, $\ln\mathcal{B}_{R, E}$ depends mainly on the ratio of prior volumes $V_E/V_R$. As discussed in Section \ref{sec:BW}, $R$ has $7N+2$ parameters and $E$ has $5N+4$. Since all \textit{BayesWave} signal models require $N\geq1$, $R$ always has more parameters than $E$, implying $V_R>V_E$. Hence for nonprecessing BBHs, the simpler model $E$ is preferred. There are no obvious correlations between $\ln\mathcal{B}_{R, E}$ and $\mathcal{O}_{R, E}$ for the nonprecessing BBHs (top left panel of Figure \ref{fig:bayesfactor}). However, as the detector network expands, $\mathcal{O}_{R, E}$ generally shifts closer to unity and $\ln\mathcal{B}_{R, E}$ becomes more negative. The accuracy of $E$ and $R$ reconstructions, and hence the similarity of their reconstructions, increases with additional detectors. As a result, their posterior distributions also become more similar, and the first and second terms of Equation \ref{eq:BRE_appx} become negligible. Hence $\ln\mathcal{B}_{R, E}$ depends more on $\ln (V_{E}/V_{R})$ and becomes more negative.

The top right panel of Figure~\ref{fig:bayesfactor} shows $\ln\mathcal{B}_{R, E}$ versus $\mathcal{O}_{R, E}$ for the precessing BBHs. While most events reside in the bottom right quadrant as with the nonprecessing BBHs, the top left quadrant ($\mathcal{O}_{R, E}<0.98$, $\ln\mathcal{B}_{R, E}>0$) is populated by 5\%, 17\% and 20\% of the HL, HLV and HLKV precessing BBHs. In fact there is a visible trend, where $\ln\mathcal{B}_{R, E}$ increases, as $\mathcal{O}_{R, E}$ decreases, implying that the evidence of $R$ for precessing BBHs increases when the reconstructions of $E$ and $R$ deviate further from one another. We also note that the color of the data points darken, as one moves from the bottom right to the top left quadrant, indicating that $\chi_{p, \rm init}$ increases. In fact, $\ln\mathcal{B}_{R, E}>0$ is generally observed in events with $\chi_{p, \rm init}\gtrsim0.5$. This is because BBHs with high initial precession are more likely to maintain precessional motion throughout the inspiral, leading to strong signal modulations that cannot be modeled by $E$, thereby strengthening the preference for $R$. However, as noted in Figure~\ref{fig:sample_overlaps}, deviations from elliptical polarization do not connect simply and predictably to $\chi_{p, \rm init}$, i.e. signals with high $\chi_{p, \rm init}$ can evolve into low-ellipticity signals, and vice versa. Therefore, we also observe some dark-colored (light-colored) data points in the bottom right (top left) quadrants, that do not follow the general $\chi_{p, \rm init}$ color trend. As quoted above, the percentage of events in the top left quadrant is highest for HLKV, followed by HLV then HL, which suggests that $\ln\mathcal{B}_{R, E}>0$ is more likely with larger detector networks. We also note that $\ln\mathcal{B}_{R, E}$ increases with the number of detectors $\mathcal{I}$. 

\begin{figure*}[t]
    \centering
    \includegraphics[width=\textwidth]{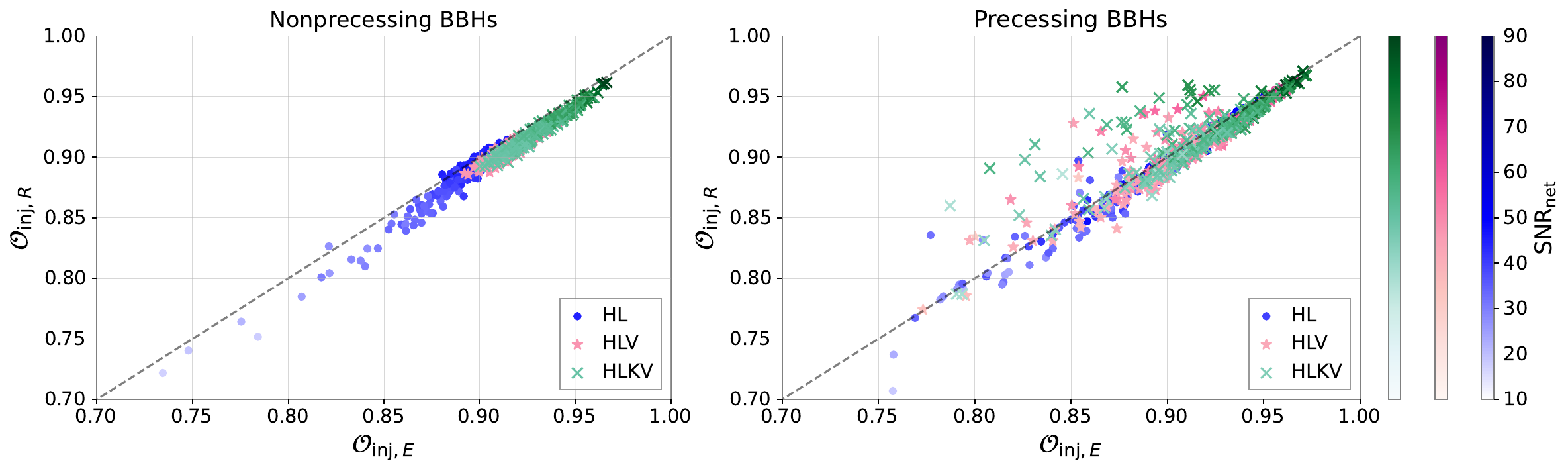}
    \caption{$\mathcal{O}_{{\rm inj}, R}$ versus $\mathcal{O}_{{\rm inj}, E}$ for the nonprecessing (left panel) and precessing (right panel) BBHs. The blue circles, pink stars and green crosses indicate HL, HLV, and HLKV events respectively. The three color bars indicate SNR$_{\rm net}$ per event for HL, HLV and HLKV in corresponding colors; this applies to both panels. The diagonal grey dashed lines in both panels indicate $\mathcal{O}_{{\rm inj}, R}=\mathcal{O}_{{\rm inj}, E}$.}
    \label{fig:overlap_compare}
\end{figure*}

In the bottom row of Figure \ref{fig:bayesfactor}, we plot $\mathcal{O}_{{\rm inj}, R}/ \mathcal{O}_{{\rm inj}, E}$ versus $\mathcal{O}_{R, E}$ to compare the reconstruction accuracy of $E$ and $R$, as their inferred waveforms deviate. As in the top row, to compare the nonprecessing and precessing BBHs, we divide the plots into four segments by placing vertical and horizontal black dashed-lines respectively at $\mathcal{O}_{R, E}=0.98$ and $\mathcal{O}_{{\rm inj}, R}/ \mathcal{O}_{{\rm inj}, E}=1$, where $E$ and $R$ reconstruct the injected signal equally well. The bottom left plot shows that nonprecessing BBHs generally fall within the bottom right quadrant, with $\mathcal{O}_{R, E}>0.98$ and $\mathcal{O}_{{\rm inj}, R}/ \mathcal{O}_{{\rm inj}, E}<1$, i.e. nonprecessing BBHs are reconstructed more accurately by $E$ than $R$. This is because nonprecessing BBHs are fundamentally elliptically polarized, so they are represented adequately by $E$. The polarization of $R$, on the other hand, is not restricted, which is disadvantageous when reconstructing fundamentally elliptical waveforms like nonprecessing BBHs, especially when the SNR is low. However SNR$_{\rm net}$ of the nonprecessing BBHs used in this study is high ($\gtrsim40$), so $E$ and $R$ perform similarly, i.e. we find $0.95<(\mathcal{O}_{{\rm inj}, R}/ \mathcal{O}_{{\rm inj}, E})<1$. Furthermore we observe a trend in the bottom left plot of Figure \ref{fig:bayesfactor}, where $\mathcal{O}_{{\rm inj}, R}/ \mathcal{O}_{{\rm inj}, E}$ decreases with $\mathcal{O}_{R, E}$. Decreasing $\mathcal{O}_{R, E}$ indicates that the waveform inferred by $R$ becomes less elliptical, resulting in lower reconstruction accuracy for the nonprecessing BBHs. We also find that $\mathcal{O}_{R, E}$ of HL is generally lower than HLV and HLKV, because HL has the lowest SNR$_{\rm net}$ amongst the three configurations. This confirms that $R$ deviates further from $E$ as the SNR decreases. In the bottom right panel of Figure \ref{fig:bayesfactor}, we see that $\mathcal{O}_{{\rm inj}, R}/ \mathcal{O}_{{\rm inj}, E}$ for the precessing BBHs follows similar trends as for the nonprecessing BBHs for $\mathcal{O}_{R, E}>0.98$. However, the trend changes for $\mathcal{O}_{R, E}<0.98$, where we observe that $\mathcal{O}_{{\rm inj}, R}/ \mathcal{O}_{{\rm inj}, E}$ increases, as $\mathcal{O}_{R, E}$ decreases. Note that the horizontal scales in the top and bottom rows of Figure \ref{fig:bayesfactor} are the same. We infer that events with $\mathcal{O}_{R, E}<0.98$ and $\mathcal{O}_{{\rm inj}, R}/ \mathcal{O}_{{\rm inj}, E}>1$ in the top left quadrant (bottom right panel) correspond to those with $\ln\mathcal{B}_{R, E}>0$ in the top right panel. In other words, when $R$ is favored by model selection, the reconstruction accuracy of $R$ is also higher than $E$. This is expected for the reasons discussed previously, that is, $R$ can reconstruct precession-induced amplitude modulations of precessing BBHs better than $E$.

Figure \ref{fig:overlap_compare} shows the event-wise comparison of $\mathcal{O}_{{\rm inj}, R}$ versus $\mathcal{O}_{{\rm inj}, E}$ for the nonprecessing (left panel) and precessing (right panel) BBHs. The information presented in this plot complements the bottom row in Figure \ref{fig:bayesfactor} by explicitly showing the reconstruction accuracy of each signal model, i.e. $\mathcal{O}_{{\rm inj}, R}$ and $\mathcal{O}_{{\rm inj}, E}$, instead of their ratio $\mathcal{O}_{{\rm inj}, R}/ \mathcal{O}_{{\rm inj}, E}$. Figure \ref{fig:overlap_compare} shows that $\mathcal{O}_{{\rm inj}, R}$ and $\mathcal{O}_{{\rm inj}, E}$ exceed 0.7 individually for both the nonprecessing and precessing BBHs, in all three detector configurations. That is, both $E$ and $R$ are able to reconstruct the precessing and nonprecessing BBHs signals reliably in general; but as discussed above, one model can outperform the other depending on the signal morphology. We find that the HLKV (green) events are reconstructed most accurately with $\mathcal{O}_{{\rm inj}, R}$ and $\mathcal{O}_{{\rm inj}, E}$ generally exceeding 0.85, followed by HLV (exceeding 0.80) and then HL (exceeding 0.70). This makes sense because HLKV events have the highest SNR$_{\rm net}$ as indicated by the color bars. This observation, where overlap increases with SNR$_{\rm net}$, is consistent with Refs. \cite{Becsy_2016, Ghonge_2020, Lee_2021}. 

In summary, we compare the multi-detector network performance of the two \textit{BayesWave} signal models, $R$ and $E$, for elliptical and nonelliptical GW burst signals. We use nonprecessing and precessing BBHs to represent the elliptical and nonelliptical signals respectively. Our analyses find that signal reconstructions by $E$ and $R$ are comparable for the nonprecessing BBHs signals, so the simpler $E$ model attracts a lower Occam penalty and is typically preferred over $R$. The same is true for most ($\gtrsim80\%$) of the precessing BBHs because, even though the polarizations of all precessing BBHs are non-elliptical to some extent, most do not deviate far enough from elliptical polarizations to require $R$ for accurate signal reconstruction. However, for precessing BBHs with more pronounced nonelliptical polarizations, the $R$ reconstruction diverges from $E$, with the former providing more accurate representations of the signal. In such cases, $R$ is preferred over $E$. This is more commonly observed in BBHs whose precession amplitude is relatively large, and in larger detector networks. From these results, we conclude that: (i) $\ln\mathcal{B}_{R, E}$ and $\mathcal{O}_{R, E}$ together can provide indications for nonelliptical polarization, but are insufficient to distinguish between purely elliptical and slightly nonelliptical signals; and (ii) $R$ can improve signal characterization of GW signals for nonelliptical polarizations, especially with expanded detector networks. 


\subsection{Demonstration with O3 data}
\label{sec:O3}

\setlength{\tabcolsep}{0pt}
\rowcolors{4}{gray!15}{white}
\begin{table*}[t!]
\renewcommand{\arraystretch}{1.4}
\begin{tabular}{p{1.5cm} p{3cm}>{\centering}p{1.7cm}>{\centering}p{1.7cm}>{\centering}p{1.8cm}>{\centering}p{2cm}>{\centering}p{1.5cm}>{\centering}p{1.5cm}>{\centering\arraybackslash}p{1.5cm}}
\hline
\multirow{2.3}{*}{LVK run}
    & \multirow{2.3}{*}{Event name}
    & \multirow{2.3}{*}{Detectors}
    & \multirow{2.3}{*}{\shortstack{Network \\ SNR}}
    & \multirow{2.3}{*}{\shortstack{$\mathcal{M}$ ($M_{\odot}$)}}
    & \multirow{2.3}{*}{$\ln\mathcal{B}_{R, E}$}
    & \multirow{2.3}{*}{$\mathcal{O}_{R, E}$}
            & \multicolumn{2}{c}{Recovered SNR}   \\
    \cline{8-9}
    & & & & & & &\textbf{$E$} & \textbf{$R$} \\ \hline\hline
        
O3a & GW190421\_213856 & HL & $10.7_{-0.4}^{+0.2}$ & $31.4_{-4.6}^{+6.0}$ & $-2.27\pm0.19$ & 0.981 & 11.1 & 11.3 \\
O3a & GW190521\_074359 & HL & $25.9_{-0.2}^{+0.1}$ & $32.8_{-2.8}^{+3.2}$ & $-0.04\pm0.20$ & 0.998 & 25.8 & 25.8 \\
O3a & GW190707\_093326 & HL & $13.1_{-0.4}^{+0.2}$ & $8.4_{-0.4}^{+0.6}$ & $-4.32\pm0.04$ & 0.999 & 10.4 & 10.0 \\
O3a & GW190408\_181802 & HLV & $14.6_{-0.3}^{+0.2}$ & $18.5_{-1.2}^{+1.9}$ & $-3.08\pm0.05$ & 0.958 & 15.1 & 14.1 \\
O3a & GW190412 & HLV & $19.8_{-0.3}^{+0.2}$ & $13.3_{-0.5}^{+0.5}$ & $-12.88\pm0.06$ & 0.952 & 19.1 & 18.2 \\
O3a & GW190503\_185404 & HLV & $12.2_{-0.4}^{+0.2}$ & $29.3_{-4.4}^{+4.5}$ & $-3.95\pm0.04$ & 0.982 & 12.0 & 11.4 \\
O3a & GW190512\_180714 & HLV & $12.7_{-0.4}^{+0.3}$ & $14.6_{-0.9}^{+1.4}$ & $-4.05\pm0.04$ & 0.991 & 11.2 & 11.2 \\
O3a & GW190513\_205428 & HLV & $12.5_{-0.4}^{+0.3}$ & $21.8_{-2.2}^{+3.8}$ & $-0.01\pm0.05$ & 0.940 & 12.8 & 12.9 \\
O3a & GW190517\_055101 & HLV & $10.8_{-0.6}^{+0.5}$ & $26.5_{-4.2}^{+4.0}$ & $-2.23\pm0.04$ & 0.982 & 10.0 & 9.7 \\
O3a & GW190519\_153544 & HLV & $15.9_{-0.3}^{+0.2}$ & $44.3_{-7.5}^{+6.8}$ & $-2.14\pm0.13$ & 1.000 & 15.4 & 15.6 \\
O3a & GW190521 & HLV & $14.3_{-0.4}^{+0.5}$ & $63.3_{-14.6}^{+19.6}$ & $-1.93\pm0.04$ & 0.997 & 15.2 & 14.8 \\
O3a & GW190602\_175927 & HLV & $13.2_{-0.3}^{+0.2}$ & $48.0_{-9.7}^{+9.5}$ & $-2.28\pm0.17$ & 0.999 & 13.2 & 13.1 \\
O3a & GW190706\_222641 & HLV & $13.4_{-0.4}^{+0.2}$ & $45.6_{-9.1}^{+13.0}$ & $-1.03\pm0.18$ & 0.988 & 12.8 & 13.0 \\
O3a & GW190727\_060333 & HLV & $11.7_{-0.5}^{+0.2}$ & $29.4_{-3.7}^{+4.6}$ & $-1.63\pm0.04$ & 0.962 & 11.9 & 12.3 \\
O3a & GW190728\_064510 & HLV & $13.1_{-0.4}^{+0.3}$ & $8.6_{-0.3}^{+0.6}$ & $-2.65\pm0.04$ & 0.839 & 8.8 & 7.6 \\
O3a & GW190814 & HLV & $25.3_{-0.2}^{+0.1}$ & $6.11_{-0.05}^{+0.06}$ & $-12.35\pm0.32$ & 0.950 & 19.5 & 19.2 \\
O3a & GW190828\_063405 & HLV & $16.5_{-0.3}^{+0.2}$ & $24.6_{-2.0}^{+3.6}$ & $-6.65\pm0.05$ & 0.963 & 17.0 & 15.6 \\
O3a & GW190828\_065509 & HLV & $10.2_{-0.5}^{+0.4}$ & $13.4_{-1.0}^{+1.4}$ & $-1.17\pm0.12$ & 0.999 & 6.5 & 6.3 \\
O3a & GW190915\_235702 & HLV & $13.1_{-0.3}^{+0.2}$ & $24.4_{-2.3}^{+3.0}$ & $-1.16\pm0.04$ & 0.990 & 12.9 & 13.6 \\
O3a & GW190630\_185205 & LV & $16.4_{-0.3}^{+0.2}$ & $25.1_{-2.1}^{+2.2}$ & $-1.72\pm0.05$ & 0.991 & 14.7 & 15.0 \\
O3a & GW190708\_232457 & LV & $13.4_{-0.3}^{+0.2}$ & $13.1_{-0.6}^{+0.9}$ & $-2.97\pm0.21$ & 0.918 & 11.8 & 10.7 \\
O3a & GW190910\_112807 & LV & $14.5_{-0.3}^{+0.2}$ & $33.5_{-4.1}^{+4.2}$ & $-1.26\pm0.17$ & 0.995 & 14.1 & 14.1 \\
\hline
   O3b & GW191109\_010717 & HL & $17.3_{-0.5}^{+0.5}$ & $47.5_{-7.5}^{+9.6}$ & $-1.91\pm0.05$ & 0.998 & 17.5 & 17.5 \\
O3b & GW191222\_033537 & HL & $12.5_{-0.3}^{+0.2}$ & $33.8_{-5.0}^{+7.1}$ & $-2.23\pm0.20$ & 0.997 & 11.7 & 11.8 \\
O3b & GW200225\_060421 & HL & $12.5_{-0.4}^{+0.3}$ & $14.2_{-1.4}^{+1.5}$ & $-2.85\pm0.19$ & 0.997 & 12.6 & 12.4 \\
O3b & GW191215\_223052 & HLV & $11.2_{-0.4}^{+0.3}$ & $18.4_{-1.7}^{+2.2}$ & $-3.78\pm0.04$ & 0.993 & 9.9 & 9.6 \\
O3b & GW200129\_065458 & HLV & $26.8_{-0.2}^{+0.2}$ & $27.2_{-2.3}^{+2.1}$ & $0.48\pm0.35$ & 0.984 & 26.9 & 27.3 \\
O3b & GW200208\_130117 & HLV & $10.8_{-0.4}^{+0.3}$ & $27.7_{-3.1}^{+3.7}$ & $-3.02\pm0.04$ & 0.980 & 9.6 & 7.9 \\
O3b & GW200219\_094415 & HLV & $10.7_{-0.5}^{+0.3}$ & $27.6_{-3.8}^{+5.6}$ & $-4.30\pm0.20$ & 0.907 & 12.0 & 10.2 \\
O3b & GW200224\_222234 & HLV & $20.0_{-0.2}^{+0.2}$ & $31.1_{-2.7}^{+3.3}$ & $-3.74\pm0.31$ & 0.997 & 20.1 & 20.3 \\
O3b & GW200311\_115853 & HLV & $17.8_{-0.2}^{+0.2}$ & $26.6_{-2.0}^{+2.4}$ & $-3.41\pm0.05$ & 0.942 & 17.4 & 17.8 \\
O3b & GW191216\_213338 & HV & $18.6_{-0.2}^{+0.2}$ & $8.33_{-0.19}^{+0.22}$ & $-4.63\pm0.39$ & 0.992 & 17.7 & 16.4 \\
 \hline
\end{tabular}
\caption{O3 GW events used in $R$ versus $E$ model selection. The columns from left to right quote: (i) the LVK observing run in which the event occurred, (ii) event name, (iii) detectors operating at time of detection, (iv) network matched-filter SNR, (v) chirp mass $\mathcal{M}$ in units of solar masses $M_\odot$, (v) log Bayes factor between $R$ and $E$, $\ln \mathcal{B}_{R,E}$, (vi) network overlap between the $R$ and $E$ signal reconstructions, $\mathcal{O}_{R, E}$, and (vii) network SNR of the signals recovered by $E$ and $R$. The network matched-filter SNR and $\mathcal{M}$ are copied directly from GWTC-2 \cite{GWTC2} (O3a events) and GWTC-3 \cite{GWTC3} (O3b events), and quote the median and the 90\% symmetric credible intervals of the Bayesian posteriors, obtained from parameter estimation pipelines. $\ln \mathcal{B}_{R,E}$ and its error margins are calculated as per Equations \ref{eq:BF_TI} and \ref{eq:BF_error}. The network matched-filter SNR in this table is not to be confused with SNR$_{\rm net}$ in the main text which denotes the injected network SNR for the simulated BBHs.}
\label{table:O3_runs}
\end{table*}

We extend the $\ln\mathcal{B}_{R, E}$ versus $\mathcal{O}_{R,E}$ analyses to real GW events from O3. The analyses of simulated BBHs in Section \ref{sec:BF_results} use Gaussian noise colored by fixed PSDs as in Figure \ref{fig:PSD}. In contrast, PSDs for real data are not known a priori, and vary temporally.  For the analysis of real events, we estimate the PSD using the \textit{BayesLine} algorithm, which uses an RJMCMC to model the PSDs as a smooth Akima spline with Lorentzians to reconstruct the narrow-band spectral features~\cite{BayesLine,BayesLineII}. For each event, we use \textit{BayesLine} to produce a posterior distribution of the on-source PSD, and use the median of that posterior distribution as the PSD used in BayesWave's likelihood calculation for both the $R$ and $E$ analyses. By using the same PSD estimate, we are able to directly compare evidences between the two models.

\textit{BayesWave} signal reconstructions are known to improve with decreasing signal duration \cite{Pannarale_BW, Ghonge_2020}; short-duration signals occupy less time-frequency volume, so \textit{BayesWave} produces more compact and faithful wavelet representations. Hence, for the $R$ versus $E$ analysis, we use a subset of O3 events that are sufficiently loud and short for \textit{BayesWave} to produce reliable reconstructions. We list the events in Table~\ref{table:O3_runs}. These events are also used for \textit{BayesWave}'s waveform consistency tests previously in GWTC-2 (O3a) \cite{GWTC2} and GWTC-3 (O3b) \cite{GWTC3}. The O3 events in Table~\ref{table:O3_runs} also span a range of chirp masses $\mathcal{M}=(m_1m_2)^{3/5}/(m_1 + m_2)^{1/5}$, where $m_1$ and $m_2$ are the primary and secondary component masses. GW signals for low-$\mathcal{M}$ mergers last longer in the LVK detector frequency bands and the components merge at higher frequencies, so longer analysis segments and higher sampling rates are required to accurately reconstruct the time-frequency structure. The opposite is true for high-$\mathcal{M}$ events.  Following the general procedure of waveform reconstructions in GWTC2~\cite{GWTC2} and GWTC3~\cite{GWTC3}, we use the same settings for the segment length, sampling rate, and lower frequency bound used in the parameter estimation analysis presented in the catalogs~\footnote{The parameter estimation configurations (as well as results) can be found via the Online Gravitational-Wave Transient Catalog: \url{https://gwosc.org/eventapi/html/GWTC/}}.

\begin{figure}
    \centering
    \includegraphics[width=.49\textwidth]{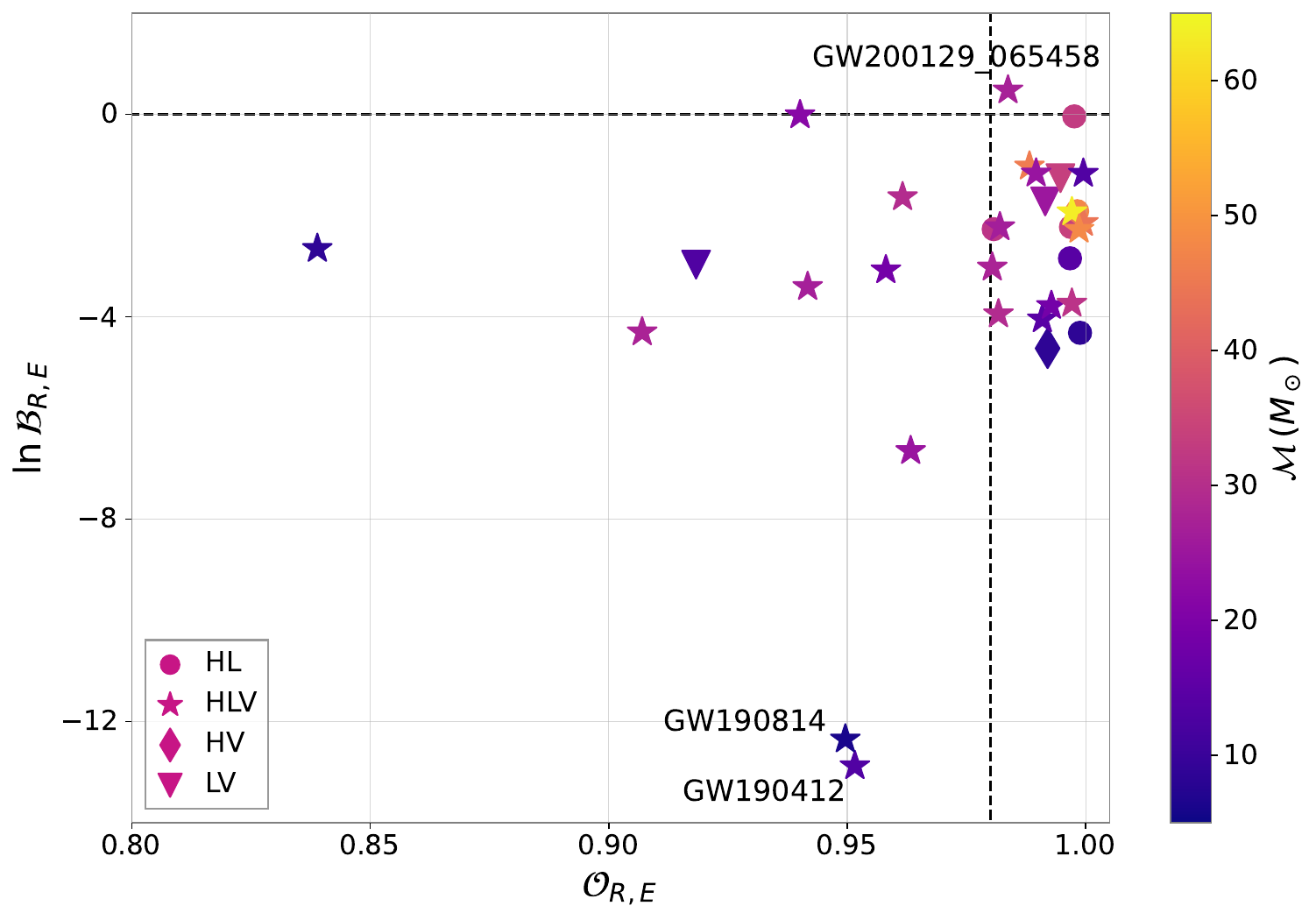}
    \caption{$\ln\mathcal{B}_{R, E}$ versus $\mathcal{O}_{R,E}$ for O3 events listed in Table \ref{table:O3_runs}. The different symbols indicate the detectors operating during the event, as shown in the legend, color coded by the chirp mass $\mathcal{M}$, in units of solar masses $M_\odot$ (see color bar). The horizontal axis is extended slightly beyond the maximum ($\mathcal{O}_{R,E}=1$) to  reduce symbol clutter at the right margin of the frame.}
    \label{fig:O3}
\end{figure}

The values of $\ln \mathcal{B}_{R,E}$ and $\mathcal{O}_{R,E}$ for each event are listed in Table \ref{table:O3_runs}. We also plot $\ln \mathcal{B}_{R,E}$ against $\mathcal{O}_{R,E}$ in Figure \ref{fig:O3} to study their mutual correlation. We divide Figure \ref{fig:O3} into the same four quadrants as in the top panels of Figure \ref{fig:bayesfactor}. We find that $\mathcal{O}_{R,E}$ for the O3 events is generally lower than for the simulated BBHs. This is because the SNRs of O3 events, as shown in the forth column of Table \ref{table:O3_runs}, are lower than for the simulated BBHs (SNR$\sim$50), resulting in larger discrepancies between $E$ and $R$. Nevertheless, we find $\mathcal{O}_{R,E}\gtrsim 0.90$ for the O3 events, indicating that $E$ and $R$ reconstructions are still comparable. Figure \ref{fig:O3} also shows that we have $\ln \mathcal{B}_{R,E}<0$ for the O3 events, and there is no obvious trend relating $\ln \mathcal{B}_{R,E}$ and $\mathcal{O}_{R,E}$. Altogether, the $\ln \mathcal{B}_{R,E}$ versus $\mathcal{O}_{R,E}$ plot of the O3 events is similar to the nonprecessing BBHs in the top left panel of Figure \ref{fig:bayesfactor}, viz. the O3 events are generally more consistent with the elliptically polarized signals characterized by $E$, compared to $R$. However, there are three events that stand out from the rest, namely GW200129$\_$065458, GW190412 and GW190814, as indicated in Figure \ref{fig:O3}. We discuss them in further detail below. 

GW200129$\_$065458 is the only event in Table \ref{table:O3_runs} with $\ln \mathcal{B}_{R,E}>0$. Ref. \cite{Hannam_2021} reports GW200129$\_$065458 as the first CBC detection with spin-precession. However, the LIGO Livingston detector experienced an instrumental glitch around the time of detection. Ref. \cite{Payne_2022} shows, using \textit{BayesWave}, that the uncertainties in glitch modelling reduces the evidence for spin-precession. Thus the presence of precession in GW200129$\_$065458 remains ambiguous. The marginally positive Bayes factor ($\ln \mathcal{B}_{R,E} = 0.48\pm0.35$) is also insufficient to make claims about the preferred polarization of GW200129$\_$065458. On the contrary, $\ln \mathcal{B}_{R,E}\approx -12$ for GW190412 and GW190814 are more negative than for other O3 events ($\ln \mathcal{B}_{R,E}\sim -5$) in Table \ref{table:O3_runs}. The color bar in Figure \ref{fig:O3} indicates that both latter events have lower $\mathcal{M}$ than most of the other O3 events and therefore last longer. Table \ref{table:O3_runs} shows that both $E$ and $R$ recover these events with high SNR ($\sim 20$). Altogether $E$ and $R$ require more wavelets to characterize GW190412 and GW190814, compared to other shorter-duration O3 signals. Recall from Sections \ref{sec:elliptical_model} and \ref{sec:relaxed_model} that $E$ and $R$ have $D_E=5N+4$ and $D_R=7N+2$ model parameters respectively. Assuming equal $N$ for both models, the difference in model dimensions beween $E$ and $R$, i.e. $D_R-D_E$, increases with $N$. Hence for GW190412 and GW190814, where $N$ is large, we have $D_R \gg D_E$. Since the $E$ and $R$ reconstructions are also similar ($\mathcal{O}_{R,E}\approx 0.95$), the simpler model $E$ is favoured over $R$. We also recognize that GW190814 and GW190412 show evidence of higher-order multipole moments \cite{Hoy_2022}. The GW signal morphology in the presence of higher-order multipoles have been shown to deviate from quadrupole-only signals \cite{IMRPhenomHM, IMRPhenomXHM}. The change in signal morphology may affect the reconstruction accuracy of $E$ and $R$, and consequently the preferred model, an interesting avenue for future work.

In summary, we perform model selection between $E$ and $R$ with O3 events. We find that $E$ and $R$ reconstructions of O3 events are comparable with $\mathcal{O}_{R,E}\gtrsim 0.90$. As with the nonprecessing BBHs in Section \ref{sec:BF_results}, $E$ is generally preferred over $R$, implying that the O3 GW signals are more consistent with elliptical polarizations than otherwise. The outlier GW200129\_065458 reveals that $\ln\mathcal{B}_{R, E}>0$ can indicate anomalous source properties, which justify further investigation with supplementary analysis pipelines.

\section{Stokes parameters with $R$}
\label{sec:GWpol}

The physical interpretation of $\epsilon$ in $E$ (as in Equation \ref{eq:E_model_cross}) is unambiguous when the underlying signal is elliptically polarized and its polarization is constant with time. On the other hand, $R$ can infer the polarization content of GW signals without constraints. In this section, we study how accurately $R$ recovers polarization content for signals with different polarizations, using the nonprecessing (elliptical) and precessing (nonelliptical) BBHs described in Section \ref{sec:injections}. We also study how the accuracy is affected by the size of detector networks using the HL, HLV and HLKV networks. 

Ref. \cite{BayesWave3} explains how Stokes parameters describe the polarization content of GW signals. We define the fraction of linear ($F_\mathrm{L}$), circular ($F_\mathrm{C}$) and total ($F_\mathrm{T}$) polarizations relative to the total signal intensity in Section \ref{sec:stokes}. In Section \ref{sec:stokes_accuracy}, we introduce a metric to quantify the accuracy between the injected and recovered $F_\mathrm{L}$, $F_\mathrm{C}$ and $F_\mathrm{T}$. We present the results for the nonprecessing and precessing BBHs using the HL, HLV, and HLKV networks in Section \ref{sec:stokes_results}.

\subsection{Stokes parameters}
\label{sec:stokes}

The polarization content of a tensor-polarized GW is described by the four Stokes parameters \cite{Jackson_ED, Romano_Cornish}
\begin{align}
        I &= |\Tilde{h}_+|^2 + |\Tilde{h}_\times|^2 \label{eq:stokes_I}\\
        Q &= |\Tilde{h}_+|^2 - |\Tilde{h}_\times|^2
        \label{eq:stokes_Q}\\
        U &= \Tilde{h}_+\Tilde{h}^\ast_\times + \Tilde{h}_\times\Tilde{h}^\ast_+
        \label{eq:stokes_U}\\
        V &= i(\Tilde{h}_+\Tilde{h}^\ast_\times - \Tilde{h}_\times\Tilde{h}^\ast_+).
        \label{eq:stokes_V}
\end{align}
Equations \ref{eq:stokes_I} and \ref{eq:stokes_V} apply to monochromatic plane waves with linearly-independent complex amplitudes $\Tilde{h}_+$ and $\Tilde{h}_\times$. A GW signal, however, is typically modulated in amplitude and frequency, i.e. $\Tilde{h}_+ = \Tilde{h}_+(f)$ and $\Tilde{h}_\times = \Tilde{h}_\times(f)$, so its Stokes parameters decompose spectrally into functions $I(f)$, $Q(f)$, $U(f)$, and $V(f)$ of the Fourier frequency $f$. By definition, $I$ is proportional to the total intensity; $V$ describes circular polarization; $Q$ and $U$ together describe linear polarization. Therefore we can measure the polarization content of GWs through the fractional circular polarization
\begin{equation}
    F_\mathrm{C}(f) = \frac{\abs{V}}{I},
\end{equation}
fractional linear polarization
\begin{equation}
    F_\mathrm{L}(f) = \frac{\sqrt{Q^2+U^2}}{I},
\end{equation}
and the total polarization
\begin{equation}
    F_\mathrm{T}(f) = \frac{\sqrt{Q^2+U^2+V^2}}{I}.
\end{equation}
Going forward we refer to the above quantities collectively as the fractional polarizations $F_\mathcal{P}$, for $\mathcal{P}\in\{\mathrm{C}, \mathrm{L}, \mathrm{T}\}$. Stokes parameters are real numbers satisfying $I^2\geq Q^2 + U^2 + V^2$, and $0\leq F_\mathcal{P}\leq1$.

\subsection{Accuracy metric: root mean squared residuals, $\mathcal{R}_{\rm RMS}$}
\label{sec:stokes_accuracy}
For $E$, the fractional polarizations $F_\mathcal{P}$ reduce to the following constants: $F_\mathrm{L}=(1-\epsilon^2)/(1+\epsilon^2)$, $F_\mathrm{C}=2\epsilon/(1+\epsilon^2)$ and $F_\mathrm{T}=1$. For $R$, $F_\mathcal{P}$ are functions of frequency. We use the root mean squared (RMS) residuals between the injected and recovered $F_\mathcal{P}$, to quantify the measurement accuracy with $R$.

\begin{figure*}[p]
    \begin{minipage}{\textwidth}
        \centering
        \includegraphics[width=\linewidth]{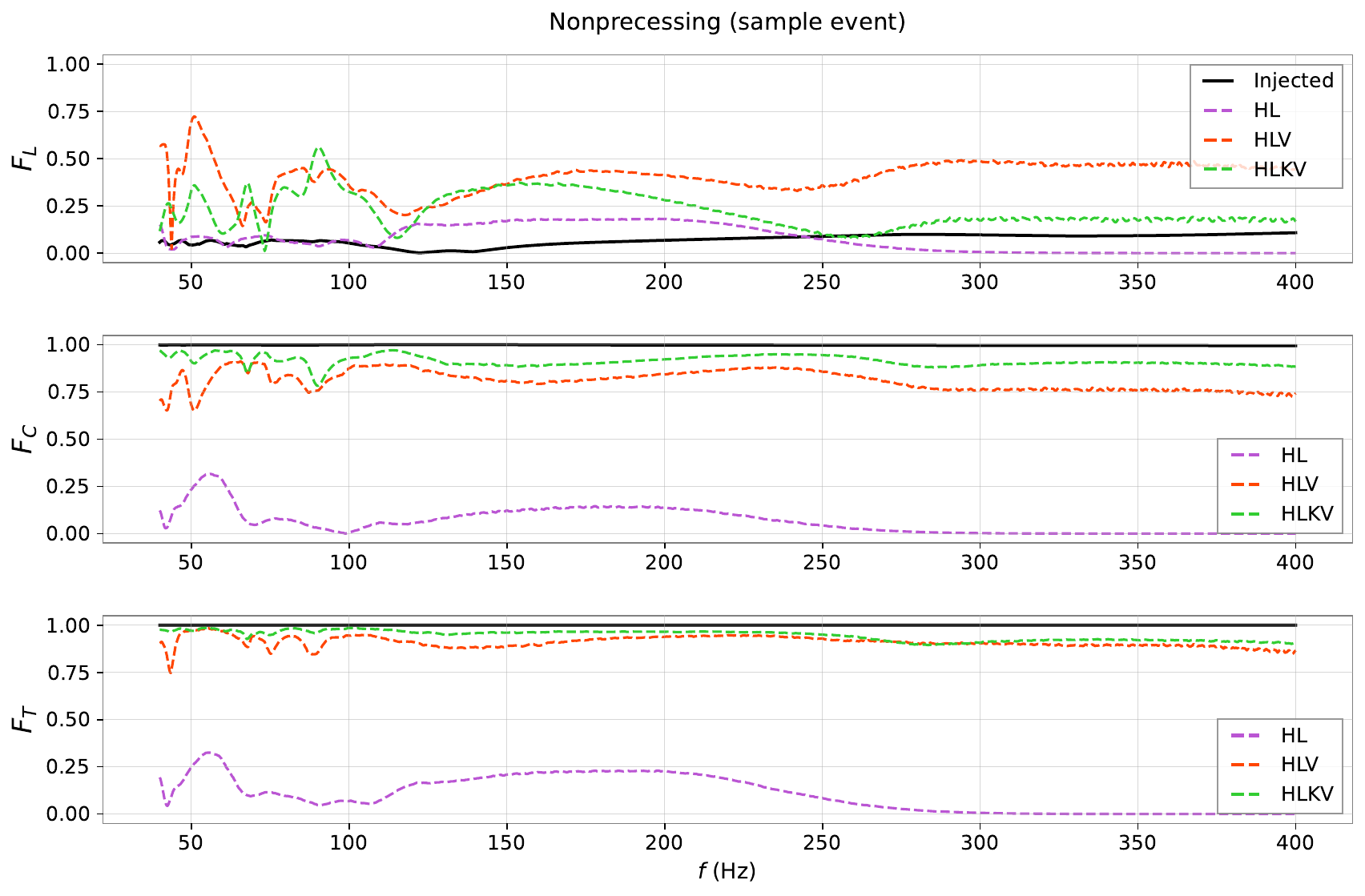}
    \end{minipage}\hfill
    \vspace{5mm}
       \begin{minipage}{\textwidth}
        \centering
        \includegraphics[width=\linewidth]{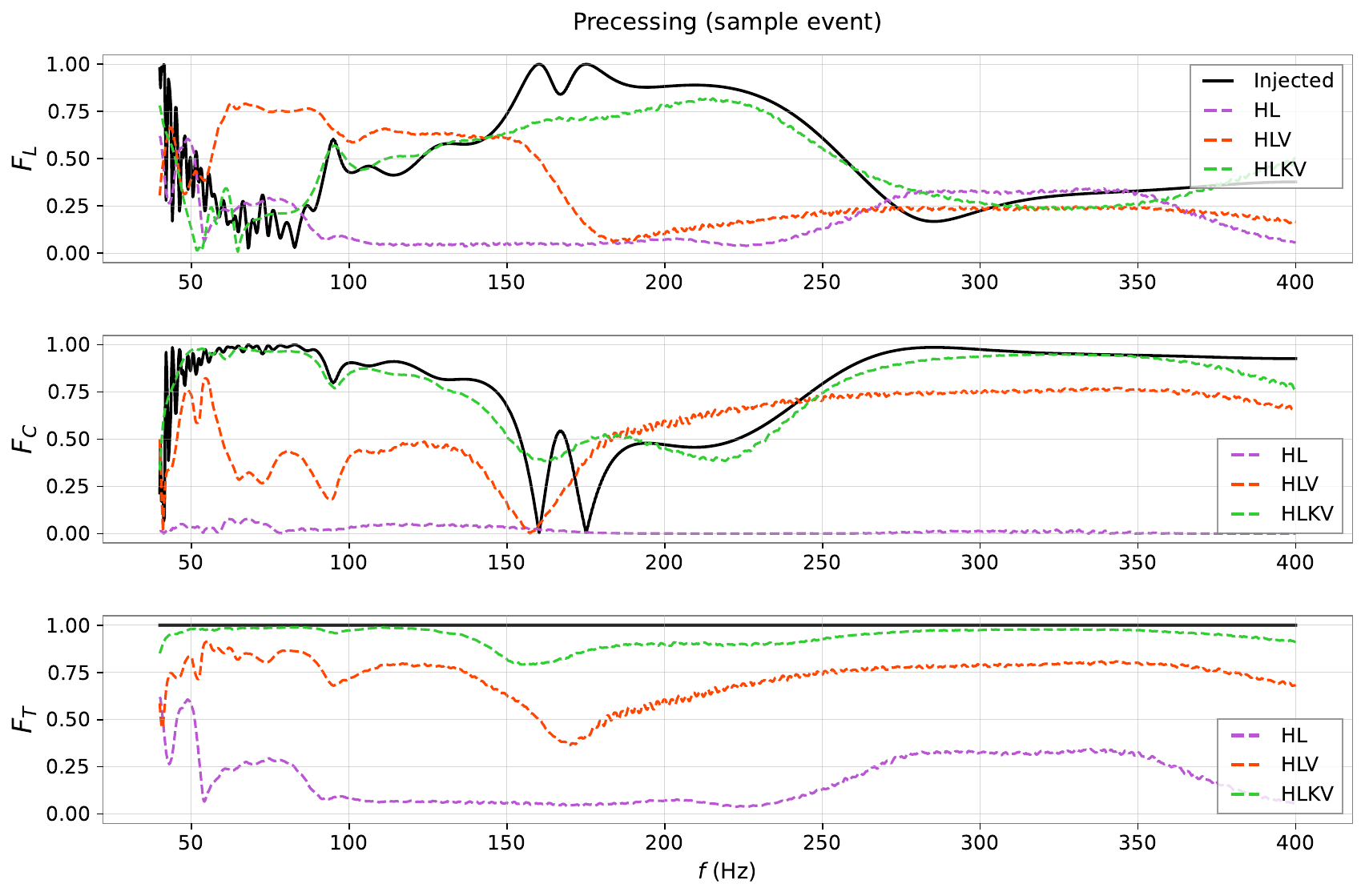}
    \end{minipage}\hfill

       \caption{Recovering Stokes parameters. Top panel: injected and recovered fractional polarizations $F_\mathrm{L}$, $F_\mathrm{C}$ and $F_\mathrm{T}$ (in order, from top to bottom) as a function of frequency $40 \, \mathrm{Hz} \leq \mathnormal{f} \leq 400 \, \mathrm{Hz}$, for a nonprecessing BBH. Bottom panel: as for top panel, but for a precessing BBH with $\chi_{p, \mathrm{init}}=0.77$. Injected $F_\mathcal{P}$ are graphed as black solid curves. Recovered $F_\mathcal{P}$ for HL, HLV, and HLKV are graphed as green, red and purple dashed curves respectively.}
    \label{fig:fracPol_event}
\end{figure*}

In \textit{BayesWave}, the  Fourier frequency $f_i$ appearing as the argument of $F_\mathcal{P}(f_i)$ is discrete, with $n$ intervals\footnote{For $n$ evenly spaced samples, i.e. $\Delta f = f_{i+1}-f_i$ for all $i\in \{1, 2,\cdots n\}$, and for a specified frequency range $f_\mathrm{min}\leq f_i \leq f_\mathrm{max}$, one can write $n=(f_\mathrm{max}-f_\mathrm{min})/\Delta f$.}. Therefore the root mean square (RMS) residuals between the injected ($\rm inj$) and recovered ($\rm rec$) $F_\mathcal{P}$ are given by
\begin{equation}
    \mathcal{R}_{\rm RMS}(F_\mathcal{P})= \sqrt{\frac{1}{n}\sum_{i=1}^{n}\left[ F_{\mathcal{P},\rm rec}(f_i) -  F_{\mathcal{P},\rm inj}(f_i) \right]^2},
    \label{eq:RMS}
\end{equation} 
with $0\leq\mathcal{R}_{\rm RMS}(F_\mathcal{P})\leq1$. Lower $\mathcal{R}_{\rm RMS}(F_\mathcal{P})$ indicates better agreement between the injected and recovered $F_\mathcal{P}$, i.e. higher measurement accuracy, and vice versa. We demonstrate this in Figure \ref{fig:fracPol_event}, which shows the injected (colored dashed lines) and recovered (solid black line) $F_\mathcal{P}$ for a representative nonprecessing (top panel) and a precessing (bottom panel) BBH from the injection sets described in Section \ref{sec:injections}. For each BBH, $F_\mathrm{L}$, $F_\mathrm{C}$ and $F_\mathrm{T}$ are plotted as functions of frequency in order, from top to bottom. To maximize the accuracy, i.e. to minimize $\mathcal{R}_{\rm RMS}(F_\mathcal{P})$, we only consider the frequency range $40 \, \mathrm{Hz} \leq \mathnormal{f} \leq 400 \, \mathrm{Hz}$ where the GW detector noise floor is lowest, as seen in Figure \ref{fig:PSD}. The two contrasting examples in Figure \ref{fig:fracPol_event} show that $F_\mathcal{P}$ recovered by HLKV (green) resembles more closely the injection (black) than HLV (red) and HL (purple). Therefore one expects $\mathcal{R}_{\rm RMS}(F_\mathcal{P})$ to be lowest for HLKV. Using $F_\mathrm{C}$ (middle row) of the precessing BBH (bottom panel) as an example, $\mathcal{R}_{\rm RMS}(F_\mathrm{C})$ of the HLKV, HLV and HL networks is given by 0.095, 0.321 and 0.810 respectively.

In Figure \ref{fig:fracPol_event}, we intentionally choose a precessing BBH with $\chi_{p, \mathrm{init}}=0.77$, i.e. substantial precession, to showcase the difference in polarization content compared to a nonprecessing BBHs, with $\chi_{p}=0$ throughout the signal. For the nonprecessing BBH (top panel), the $F_\mathcal{P}$ displays little or no frequency dependence. We find $F_\mathrm{C}>F_\mathrm{L}$ throughout the range $40 \, \mathrm{Hz} \leq \mathnormal{f} \leq 400 \, \mathrm{Hz}$, implying that the signal is more circularly polarized than linearly polarized. In contrast, for the precessing BBH example (bottom panel), $F_\mathrm{L}$ and $F_\mathrm{C}$ fluctuate with frequency, and the linear and circular polarizations dominate at different frequencies. We find $F_\mathrm{T}=1$ for both the nonprecessing and precessing examples, indicating that the signals are completely polarized at all $f$.

\subsection{Nonprecessing and precessing BBH}
\label{sec:stokes_results}

\begin{figure*}
    \begin{minipage}{\textwidth}
        \centering
        \includegraphics[width=\linewidth]{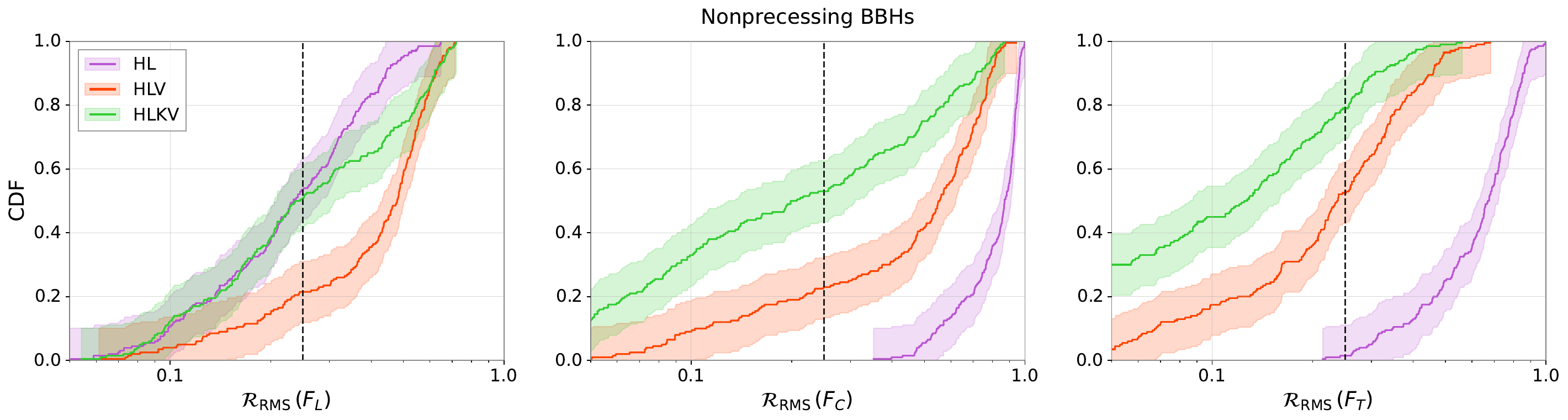}
    \end{minipage}\hfill
    \vspace{5mm}
       \begin{minipage}{\textwidth}
        \centering
        \includegraphics[width=\linewidth]{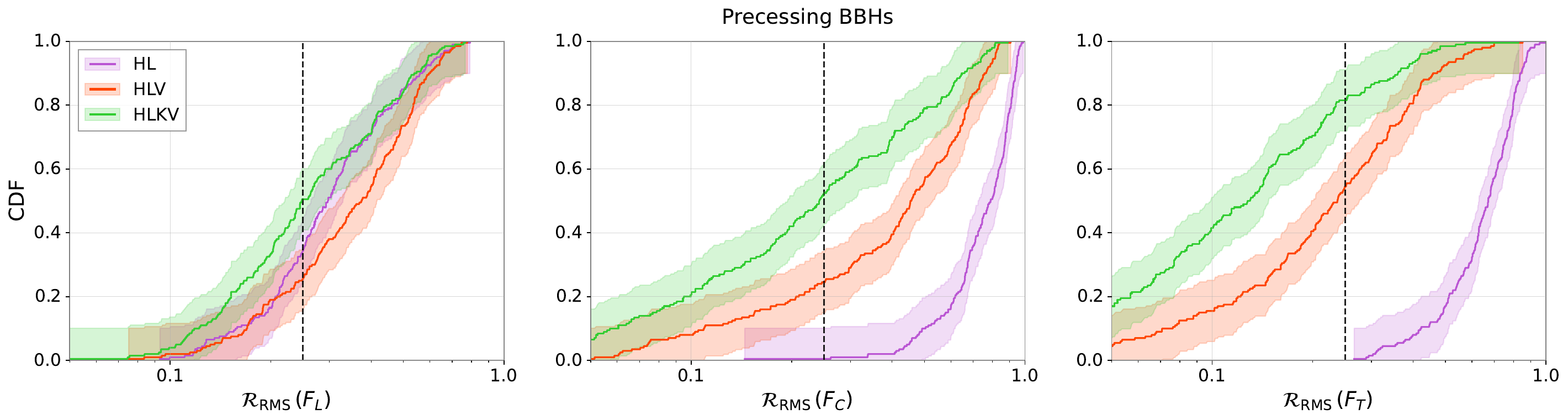}
    \end{minipage}\hfill
       \caption{Overall perspective on accuracy: cumulative distribution functions (CDFs) of Stokes parameter root mean square residuals $\mathcal{R}_{\rm RMS}(F_\mathcal{P})$. The CDFs for $\mathcal{R}_{\rm RMS}(F_\mathrm{L})$, $\mathcal{R}_{\rm RMS}(F_\mathrm{C})$ and $\mathcal{R}_{\rm RMS}(F_\mathrm{T})$ are shown in the left, middle and right columns respectively. The top and bottom rows display the CDFs for nonprecessing BBHs and precessing BBHs respectively. The CDFs for HL, HLV and HLKV are shown in purple, red and green respectively; the shaded region in corresponding colors brackets the $2$-$\sigma$ confidence interval (see Appendix \ref{sec:DKW}). The vertical dashed lines indicate $\mathcal{R}_{\rm RMS}(F_\mathcal{P})=0.25$. Higher curves imply higher measurement accuracy.}
    \label{fig:fracPol_CDF}
\end{figure*}

Figure \ref{fig:fracPol_CDF} shows the cumulative distribution functions (CDFs) of $\mathcal{R}_{\rm RMS}(F_\mathrm{L})$, $\mathcal{R}_{\rm RMS}(F_\mathrm{C})$ and $\mathcal{R}_{\rm RMS}(F_\mathrm{T})$ in order from left to right. The top (bottom) row shows the CDFs for the nonprecessing (precessing) BBHs, and CDFs for the HL, HLV and HLKV networks are shown in purple, red and green respectively. The shaded regions in corresponding colors indicate the $2$-$\sigma$ confidence intervals derived using the Dvoretzky-Kiefer-Wolfowitz inequality as described in Appendix \ref{sec:DKW}. We set a practical accuracy threshold of $\mathcal{R}_{\rm RMS}(F_\mathcal{P})=0.25$, and we only compare the CDFs for $\mathcal{R}_{\rm RMS}(F_\mathcal{P})\leq0.25$, i.e. to the left of the vertical dashed lines, in each panel of Figure \ref{fig:fracPol_CDF}. Higher CDF curves imply lower discrepancies between the injected and recovered $F_\mathcal{P}$, that is, $F_\mathcal{P}$ is recovered more accurately.

We first discuss the CDFs of $\mathcal{R}_{\rm RMS}(F_\mathrm{L})$, in the left column of Figure \ref{fig:fracPol_CDF}. For the nonprecessing BBHs (top left panel), the HL CDF at $\mathcal{R}_{\rm RMS}(F_\mathrm{L})\leq 0.25$ is 0.54. For precessing BBHs (bottom left panel), it is 0.34. In other words, the $F_\mathrm{L}$ measurement accuracy of HL is higher for nonprecessing BBHs than for precessing BBHs. The contrast is due to the different linear polarization content, which we justify as follows. For a given polarization fraction $F_\mathcal{P}$, we write its average across the frequency range $40 \, \mathrm{Hz} \leq \mathnormal{f_i} \leq 400 \, \mathrm{Hz}$ as
\begin{equation}
    \overline{F_\mathcal{P}} = \frac{\sum_{i=1}^n F_{\mathcal{P}, \mathrm{inj}}(f_i)}{n}.
\label{eq:avgPol}
\end{equation}
The median $\overline{F_\mathrm{L}}$ for the nonprecessing and precessing BBHs are 0.06 and 0.16 respectively. In Figure \ref{fig:fracPol_event}, one can see that $F_\mathcal{P}$ recovered by HL is generally close to zero, regardless of the injected $F_\mathcal{P}$. Since $\overline{F_\mathrm{L}}$ of the nonprecessing BBHs is generally lower than the precessing BBHs, HL recovers $F_\mathrm{L}$ of nonprecessing BBHs more accurately by coincidence. The HLV $\mathcal{R}_{\rm RMS}(F_\mathrm{L})$ CDFs, on the other hand, are quantitatively comparable between the nonprecessing and precessing BBHs; at $\mathcal{R}_{\rm RMS}(F_\mathrm{L})=0.25$, the HLV CDFs are respectively 0.22 and 0.26 for the nonprecessing and precessing BBHs. The same is true for HLKV, where the CDFs at $\mathcal{R}_{\rm RMS}(F_\mathrm{L})=0.25$ are the same (0.51) for both BBHs. These observations suggest that networks with three or more detectors measure low amplitude linear polarizations more consistently compared to a two-detector network, i.e. the measurement accuracy is not affected by signal morphology. Furthermore, the CDFs of HLKV (green) are higher than HLV (red) for all BBHs in the range $\mathcal{R}_{\rm RMS}(F_\mathcal{P})\leq0.25$, implying that larger detector networks recover $F_\mathrm{L}$ more accurately.

Next we discuss $\mathcal{R}_{\rm RMS}(F_\mathrm{C})$ in the middle column of Figure \ref{fig:fracPol_CDF}. For nonprecessing BBHs (top middle panel), the CDF of HLKV is highest in the practically useful range $\mathcal{R}_{\rm RMS}(F_\mathrm{C})\leq0.25$, followed by HLV and then HL. The confidence intervals overlap marginally, suggesting that the differences in the CDFs are statistically significant. The same is true for precessing BBHs (bottom middle). At $\mathcal{R}_{\rm RMS}(F_\mathrm{C})=0.25$ the CDF of HL, HLV and HLKV is 0.0, 0.23, 0.53 respectively for the nonprecessing BBHs and 0.01, 0.25 and 0.52 for the precessing BBHs. These values suggest that the CDFs of the nonprecessing and precessing BBHs are quantitatively similar for the same detector networks. In other words, the polarization of the signal does not affect the measurement accuracy, if the same detector configuration is used. Additionally, the CDFs increase with the number of detectors, which suggests that larger detector networks recover $F_\mathrm{C}$ more accurately. We find that for HL, the $\mathcal{R}_{\rm RMS}(F_\mathrm{C})$ CDF is generally lower than $\mathcal{R}_{\rm RMS}(F_\mathrm{L})$, suggesting that HL recovers $\overline{F_\mathrm{C}}$ less accurately than $\overline{F_\mathrm{L}}$. The median $\overline{F_\mathrm{C}}$ of the nonprecessing and precessing BBHs are 1.0 and 0.97 respectively, higher then $\overline{F_\mathrm{L}}$. This confirms that HL performs relatively poorly in recovering nontrivial fractional polarizations $F_\mathcal{P}$. The same arguments apply to $\mathcal{R}_{\rm RMS}(F_\mathrm{T})$, in the right column of Figure \ref{fig:fracPol_CDF}. 

Altogether, we find that \textit{BayesWave}'s signal model $R$ recovers polarization content of GW bursts more accurately as the detector network expands. We also find that $F_\mathcal{P}$ of nonprecessing and precessing BBHs are recovered with comparable accuracy using networks with three or more detectors, i.e. HLV and HLKV. This observation suggests that the underlying signal morphology (e.g. precessing or nonprecessing) does not affect the performance of $R$ in recovering polarization content, when the detector network is sufficiently large.

\section{Conclusions}
\label{sec:conclusion}

The \textit{BayesWave} algorithm offers two tensor-polarized signal models: the elliptical polarization model $E$ and the relaxed polarization model $R$. This paper studies two aspects of their performance: (i) the signal characterization by $R$ compared to $E$ in Section \ref{sec:RvsE}, and (ii) the accuracy of $R$ in measuring GW burst polarizations via Stokes parameters in Section \ref{sec:GWpol}. The expanding global detector network enables better signal characterization and hence more accurate measurements of GW polarizations, so we present multi-detector network comparisons between the HL (two-detector), HLV (three-detector) and HLKV (four-detector) networks to study how the network size impacts the performance of $E$ and $R$.

Ref. \cite{BayesWave3} finds quanlitatively that $R$ reconstructs GW burst signals with time-varying polarization content better than $E$, using precessing BBHs as an example. However, model selection does not depend entirely on reconstruction accuracy; $R$ has more model parameters than $E$ and incurs an Occam penalty in its Bayesian model evidence. In Section \ref{sec:RvsE}, we conduct a quantitative study of how the Bayes factor between $R$ and $E$ ($\ln\mathcal{B}_{R,E}$) is related to the discrepancy of reconstructed waveforms between $R$ and $E$ ($\mathcal{O}_{R,E}$). The analysis uses two sets of simulated BBHs, nonprecessing and precessing, which produce elliptical and nonelliptical GW polarizations respectively. We find that $E$ and $R$ reconstruct GW signals from nonprecessing BBHs comparably well and therefore $E$ is preferred over $R$ for its simplicity. We also note that $\mathcal{O}_{R,E}$ for nonprecessing BBHs is generally higher with larger detector networks, because additional detectors enhances SNR and hence improves the reconstruction accuracy for both models. This results in a stronger preference for $E$ in larger detector networks like HLV and HLKV. Similar trends apply for precessing BBH, but we find that $R$ is preferred in some cases, especially for $\chi_{p, \rm init}\gtrsim0.5$. High $\chi_{p, \mathrm{init}}$ indicates high initial precession, where the GW signals are more likely to possess time-varying (nonelliptical) polarization, whereupon they are better modelled by $R$. As a result, signal reconstructions of $E$ and $R$ generally show larger discrepancies for high $\chi_{p, \mathrm{init}}$ events, where $E$ is less consistent with the data, and the preference for $R$ increases. For the precessing BBHs, we also find that the preference for $R$ increases with larger detector networks, suggesting that additional detectors enable $R$ to characterize nonelliptical polarizations more accurately. Altogether the results imply empirically, that $\ln\mathcal{B}_{R, E}$ and $\mathcal{O}_{R, E}$ together indicate reliably when a GW signal deviates from elliptical polarization. However, the simulated BBHs are injected at unrealistically high SNR for the proof-of-concept study in Section \ref{sec:BF_results}. Therefore Section \ref{sec:O3} repeats the study using 32 real GW events from O3. The study finds that $E$ is generally preferred over $R$, suggesting perhaps that $\chi_{p,\rm init}$ is relatively low in most of the O3 events. We reiterate that $\ln\mathcal{B}_{R, E}$ and $\mathcal{O}_{R, E}$ cannot make definitive claims about the nature of the O3 events, but they can motivate follow-up analyses for events with potentially interesting properties.

Even though model selection generally prefers $E$ over $R$, $E$ cannot be used to measure the polarization content of generic GW bursts because it assumes elliptical polarization regardless of the true underlying polarization structure. $R$, on the other hand, is not restricted to a fixed polarization structure. In Section \ref{sec:GWpol}, we demonstrate how $R$ measures generic polarization content, using the same nonprecessing and precessing BBHs as above. We use the frequency-dependent fractional circular ($F_\mathrm{C}$), fractional linear ($F_\mathrm{L}$) and total ($F_\mathrm{T}$) polarizations derived from the Stokes parameters to quantify polarization as a function of frequency $f$ for polychromatic GW signals. For $F_\mathcal{P}$ with $\mathcal{P}\in\{\mathrm{C, L, T}\}$, we use the root mean squared residuals $\mathcal{R}_{\rm RMS}(F_\mathcal{P})$ to quantify the accuracy of $R$. As expected, we find that larger detector networks measure polarization more accurately; the overall $\mathcal{R}_{\rm RMS}(F_\mathcal{P})$ CDFs of HLKV is higher than HLV by a factor of $\sim 2$. We also find the $\mathcal{R}_{\rm RMS}(F_\mathcal{P})$ CDFs for nonprecessing and precessing BBHs are approximately equal, when one observes with either HLV or HLKV, i.e. the accuracy of polarization measurement is independent of signal morphology, when the detector network is sufficiently large. The inferior consistency and accuracy of the smaller HL network is attributed to degeneracies stemming from detector alignment. Avenues for future work include quantifying polarization measurements for other two-detector configurations, e.g. HV or LV, and comparing their performances to HLV and HLKV. We also recommend looking into reconstructions of unpolarized signals beyond BBHs, e.g. CCSNe and generic white noise bursts, to understand more fully the performance of $R$. 

How do the results of this paper impact future tests of GR? Various analyses have been conducted to probe the polarization content of CBC signals by comparing GR models against alternative theories of gravity \cite{GRtest_GWTC1, GRtest_GWTC2, GRtest_GWTC3, Hiroki_1, Hiroki_2, Isi_2017}. Supplementary approaches are based on generalizing the parameterized post-Einsteinian \cite{Yunes_2009, Cornish_2011, PhysRevD.86.022004} and null-stream frameworks\footnote{Null streams are linear combinations of detector outputs which are insensitive to tensor polarizations for a specified sky location. Null-stream fluctuations obey a chi-squared distribution, if the sky location is known exactly. \cite{Chatziioannou_2017, Hagihara_2019, Pang_2020}.} \cite{Yekta_1989}. The null-stream approach does not rely on waveform templates and has been implemented as a consistency check when analyzing binary black hole (BBH) mergers in GWTC-2 and -3 \cite{GRtest_GWTC2, GRtest_GWTC3}. However, this method requires exact knowledge of the source location and forfeits phase information \cite{GRtest_GWTC2}. By further generalizing $R$ to allow up to six independent polarization modes, \textit{BayesWave} offers a model-independent framework to probe polarizations of GW bursts without the above limitations \cite{Chatziioannou_2021}. In principle, one can compare the Bayes factor between the non-GR (non-tensor) and GR (tensor) polarization models. However, model evidences are sensitive to the priors. That is, one can arbitrarily increase the prior range of a model to decrease preference for the model, or vice versa. \textit{BayesWave} generally uses flat priors for both the intrinsic and extrinsic model parameters \cite{BayesWave3}. Since $R$ and $E$ are both tensor-polarized, i.e. they both have exactly two polarization modes, they share similar parameters and hence similar priors. In contrast, if both the non-GR and GR signal models use flat priors, the larger prior volume of the non-GR model with six polarizations attracts a larger Occam penalty, biasing the model selection towards the GR models. As the global detector expands and more GWs are detected, the priors of nontensorial polarizations are expected to become better constrained. Therefore the methods presented in Section \ref{sec:RvsE} of this paper can be adapted to distinguish between GR and non-GR polarizations in the future.

\section*{Acknowledgements}

This material is based upon work supported by NSF's LIGO Laboratory which is a major facility fully funded by the National Science Foundation. Parts of this research were conducted by the Australian Research Council Centre of Excellence for Gravitational Wave Discovery (OzGrav), through project numbers CE170100004 and CE230100016. The authors are grateful for computational resources provided by the LIGO Laboratory and supported by National Science Foundation Grants PHY-0757058 and PHY-0823459. YS. C. Lee. is supported by a Melbourne Research Scholarship and the Jean Laby Women in Physics Travel Award. MM gratefully acknowledges the NSF for support from Grant PHY-2110481. SD was supported by the Letson Summer Internship Award. The authors also thank Katerina Chatziioannou, Neil Cornish and Megan Arogeti for their helpful comments.

\appendix

\section{Bayesian model evidence via thermodynamic integration}
\label{sec:TI}

\textit{BayesWave} uses parallel tempering \cite{Parallel_tempering} to improve convergence of the RJMCMC to the target (posterior) distribution. A parallel-tempered MCMC allows multiple Markov chains with different `temperatures' ($T$) to run in parallel, while occasionally letting the chains exchange positions to improve coverage of the parameter space. Let $p(\mathbf{d}|\boldsymbol{\theta}, \mathcal{H}) $ denote the likelihood of the data $\mathbf{d}$ for a given model $\mathcal{H}$ parameterized by $\boldsymbol{\theta}$. The temperature modifies the likelihood according to $p(\mathbf{d}|\boldsymbol{\theta}, \mathcal{H})\mapsto p(\mathbf{d}|\boldsymbol{\theta}, \mathcal{H})^{1/T}$, viz. high $T$ chains survey more of the prior volume, whereas low $T$ chains explore regions surrounding the posterior distribution. In addition to improving convergence, the chains can be used to directly calculate the Bayesian model evidence using thermodynamic integration \cite{BW_TI}. The procedure is detailed below.

The model evidence is the likelihood of producing the data $\mathbf{d}$ given model $\mathcal{H}$ parameterized by $\boldsymbol{\theta}$, which by definition is the likelihood of $\mathcal{H}$ marginalized over the domain of $\boldsymbol{\theta}$, i.e.
\begin{equation}
    p(\mathbf{d}|\mathcal{H}) = \int d\boldsymbol{\theta} \, \, p(\boldsymbol{\theta}|\mathcal{H})p(\mathbf{d}|\boldsymbol{\theta}, \mathcal{H}).
\end{equation}
$p(\boldsymbol{\theta}|\mathcal{H})$ is the prior distribution of $\boldsymbol{\theta}$. By analogy, one can compute the evidence  for $\mathcal{H}$ at a given temperature $T=1/\beta$:
\begin{equation}
    Z(\beta) = \int d\boldsymbol{\theta} \, \,  p(\boldsymbol{\theta}|\mathcal{H})p(\mathbf{d}|\boldsymbol{\theta}, \mathcal{H})^{\beta},
    \label{eq:partition}
\end{equation}
with $0\leq\beta\leq1$. $Z(\beta)$ is equivalent to a partition function in a physical system described by thermodynamic variables. Since the prior is independent of $\beta$, Equation \ref{eq:partition} can be rewritten as
\begin{equation}
\begin{split}
\frac{d}{d\beta} \ln Z(\beta) &= \int d\boldsymbol{\theta} \, \, \left[\frac{p(\boldsymbol{\theta}|\mathcal{H}) p(\mathbf{d}|\boldsymbol{\theta}, \mathcal{H})^{\beta}}{Z(\beta)} \right]\ln p(\mathbf{d}|\boldsymbol{\theta}, \mathcal{H}),
\end{split}
\label{eq:expval_likelihood}
\end{equation}
using the chain rule 
\begin{equation}
\dv{\beta} \ln f(\beta) = \frac{1}{f(\beta)}\dv{f}{\beta}.
\end{equation}
According to Bayes' Theorem, the terms enclosed within the square brackets in Equation \ref{eq:expval_likelihood} collectively represent the posterior probability $p_\beta(\boldsymbol{\theta}|\mathbf{d}, \mathcal{H})$ of $\boldsymbol{\theta}$ for the chain at temperature $T=1/\beta$. Therefore the right-hand side of Equation \ref{eq:expval_likelihood} is, by definition, the expectation value of the log likelihood $\expval{\ln p(\mathbf{d}|\boldsymbol{\theta}, \mathcal{H})}_\beta$ for the corresponding $\beta$. This quantity can be evaluated directly from the RJMCMC chain \cite{Gregory2005}:
\begin{equation}
    \expval{\ln p(\mathbf{d}|\boldsymbol{\theta}, \mathcal{H})}_\beta = \frac{1}{\mu}\sum_{i=1}^\mu\ln p(\mathbf{d}|\boldsymbol{\theta}_{i,\beta}, \mathcal{H}).
\end{equation}
where $\mu$ denotes the number of samples in the chain after the burn-in period and $\{\boldsymbol{\theta}_{i,\beta}\}$ represents the set of samples for the $T=1/\beta$, for $i\in\{1, 2, \cdots, \mu\}$.
 
The model evidence $p(\mathbf{d}|\mathcal{H})$ can be interpreted similarly to the Helmholtz free energy of a canonical physical ensemble \cite{TI}, i.e. the logarithm of the partition function $\ln Z(\beta)$. This can be obtained by integrating Equation \ref{eq:expval_likelihood} over $\beta$:
\begin{align}
          \ln p(\mathbf{d}|\mathcal{H}) &= \ln Z(\beta) \\
        &= \int_0^1 d\beta \, \, \expval{\ln p(\mathbf{d}|\boldsymbol{\theta}, \mathcal{H})}_\beta.           \label{eq:TD_integration}
\end{align}
Due to the finite number of parallel tempering chains, the integral \ref{eq:TD_integration} is evaluated as a discrete sum in practice and is subject to discretization error. Additionally, $\expval{\ln p(\mathbf{d}|\boldsymbol{\theta}, \mathcal{H})}_\beta$ is estimated using the RJMCMC marginalization of model parameters and therefore contain statistical errors. It is, however, challenging to estimate these errors analytically. In \textit{BayesWave}, the integral of Equation \ref{eq:TD_integration} is computed at each RJMCMC iteration, which produces a posterior distribution of $\ln p(\mathbf{d}|\mathcal{H})$. The expectation value $E\left[\ln p(\mathbf{d}|\mathcal{H})\right]$ and variance $V\left[\ln p(\mathbf{d}|\mathcal{H})\right]$ of the posterior provide the central value and error estimate of $\ln p(\mathbf{d}|\mathcal{H})$ respectively. That is, the log Bayes factor between two models $\mathcal{H}_1$ and $\mathcal{H}_2$ is given by
\begin{equation}
    \ln \mathcal{B}_{1,2} = E\left[\ln p(\mathbf{d}|\mathcal{H}_1)\right]- E\left[\ln p(\mathbf{d}|\mathcal{H}_2)\right]
\label{eq:BF_TI}
\end{equation}
and has an error margin
\begin{equation}
    \Delta \ln \mathcal{B}_{1,2} = \sqrt{V\left[\ln p(\mathbf{d}|\mathcal{H}_1)\right] + V\left[\ln p(\mathbf{d}|\mathcal{H}_2)\right]}.
\label{eq:BF_error}
\end{equation}

\section{Precession spin parameter, $\chi_p$}

\label{sec:chi_p}

BBHs are generally characterized by eight intrinsic physical parameters: the component masses ($m_1$, $m_2$) and the spin angular momentum vectors ($\vb*{S}_1$, $\vb*{S}_2$) \cite{Pop_GWTC}. The indices $1$ and $2$ denote the primary and secondary black holes respectively. For brevity, we write $\vb*{S}_\mathcal{C}$ for $\mathcal{C}=1,2$ when discussing properties that are relevant to both $\vb*{S}_1$ and $\vb*{S}_2$. $\vb*{S}_\mathcal{C}$ can be decomposed into spin components that are parallel ($\vb*{S}_{\mathcal{C}, \parallel})$ and orthogonal ($\vb*{S}_{\mathcal{C}, \perp}$) to the binary orbital angular momentum vector $\vb*{L}$:
\begin{equation}
    \vb*{S}_\mathcal{C} = \vb*{S}_{\mathcal{C}, \parallel} + \vb*{S}_{\mathcal{C}, \perp}.
    \label{eq:spin_components}
\end{equation}
$\vb*{S}_{\mathcal{C}, \perp}$ is otherwise known as the in-plane spin and can be used to approximate the effective precession of a binary system through a single precession spin parameter $\chi_p$ \cite{Schmidt_2015}. We summarize its derivation as follows. 

The leading order of the post-Newtonian precession equation is given by \cite{Apostolatos_PN}
\begin{equation}
\begin{split}
    \dot{\vb*{L}} = \frac{1}{r^2}\left( A_1\vb*{S}_{1} + A_1\vb*{S}_{2}\right)\times \vb*{L},
    \label{eq:PN_precess}
\end{split}    
\end{equation}
with $A_1=2+3/(2q)$, $A_2=2+3q/2$, and $q=m_1/m_2 \geq 1$, where $r$ denotes the orbital separation. Since one has $\vb*{S}_{c, \parallel}\times \vb*{L}=0$ by definition, we can rewrite Equation \ref{eq:PN_precess} as
\begin{equation}
\dot{\vb*{L}}=\frac{1}{r^2}\left( A_1\vb*{S}_{1, \perp} + A_1\vb*{S}_{2, \perp}\right)\times \vb*{L},
    \label{eq:PN_precession}
\end{equation}
to indicate explicitly that the time evolution of $\vb*{L}$ in a precessing system is only driven by $\vb*{S}_{\mathcal{C}, \perp}$. In a precessing system, the angle between $\vb*{S}_{\mathcal{C}, \perp}$ and $\vb*{L}$ as well as its magnitude $|\vb*{S}_{\mathcal{C}, \perp}|=S_{\mathcal{C}, \perp}$ change with time. Figure 5 of Ref. \cite{Schmidt_2015} shows that the magnitude $|A_1\vb{S}_{1, \perp} + A_1\vb*{S}_{2, \perp}|$ of Equation \ref{eq:PN_precession} oscillates about a mean value which is consistent over multiple precession cycles. The oscillation amplitudes is small, so precise modelling of this quantity is not necessary for accurate description of the waveform. Instead, the overall in-plane spin magnitudes in Equation \ref{eq:PN_precession} can be approximated by a single parameter
\begin{align}
        S_p &= \frac{1}{2}\left[ (A_1S_{1, \perp} + A_2S_{2, \perp}) + \abs{A_1S_{1, \perp} - A_2S_{2, \perp}}\right]\\
    &= \max (A_1S_{1, \perp}, A_2S_{2, \perp}),
\end{align}
which effectively represents the mean oscillation amplitude of Equation \ref{eq:PN_precession}. By definition, $S_p$ is the average between the maximum and minimum in-plane spin contributions of $\vb*{S}_{1, \perp}$ and $\vb*{S}_{2, \perp}$, that is when they are parallel and anti-parallel respectively. 

The spin angular momentum $\vb*{S}_\mathcal{C}$ is conventionally referenced using its dimensionless counterpart $\vb*{\chi}_\mathcal{C}=\vb*{S}_\mathcal{C}/m_\mathcal{C}^2$, with $0\leq \abs{\vb*{\chi}_\mathcal{C}}\leq 1$. By analogy one can define a dimensionless precession spin parameter
\begin{equation}
    \chi_p =\frac{S_p}{A_1m_1^2}.
    \label{eq:chi_p}
\end{equation}
Note that the primary black hole appears in the denominator of Equation \ref{eq:chi_p}. This is done because $S_p$ typically reduces to $A_1S_{1,\perp}$, because the primary black hole spin tends to dominate, as $q$ increases \cite{XPHM4}. For an interpretation relevant to similar-mass binaries, we refer the reader to Section III of Ref. \cite{Schmidt_2015} for further details.

\section{Confidence intervals of cumulative distribution functions} 
\label{sec:DKW}

Dvoretzky-Kiefer-Wolfowitz (DKW) derived an inequality which can be used to compute confidence intervals for empirically determined cumulative distribution functions (CDF) \cite{DKW}. For an empirical CDF, $P(x)$, with $n$ data points, the interval which contains the true distribution $P_\text{true}(x)$ with a probability $\mathcal{L}$ is given by
\begin{equation}
    P_n(x) - \varepsilon \leq P_\text{true}(x) \leq P_n(x) + \varepsilon,
\end{equation}
with
\begin{equation}
    \varepsilon = \sqrt{\frac{\ln\left(\frac{2}{1-\mathcal{L}}\right)}{2n}}.
\end{equation}
The $2$-$\sigma$ confidence intervals in Figure \ref{fig:fracPol_CDF} of the main text are calculated using $\mathcal{L}=0.95$.

\bibliography{citations}

\end{document}